\documentclass{aastex63}
\usepackage{rotating}
\usepackage{multirow}

\accepted{January 2, 2019}
\submitjournal{ApJ}

\shorttitle{The Sharpest view of the molecular gas in the Central Region of NGC 6240}
\shortauthors{Treister et al.}

\turnoffedit

\begin{document}

\title{The Molecular Gas in the NGC 6240 Merging Galaxy System at the Highest Spatial Resolution}

\correspondingauthor{Ezequiel Treister}
\email{etreiste@astro.puc.cl}

\author[0000-0001-7568-6412]{Ezequiel Treister}
\affiliation{Instituto de Astrof{\'i}sica, Facultad de F{\'i}sica, Pontificia Universidad Cat{\'o}lica de Chile, Casilla 306, Santiago 22, Chile}

\author[0000-0002-2985-7994]{Hugo Messias}
\affiliation{Joint ALMA Observatory, Alonso de C\'{o}rdova 3107, Vitacura, Santiago, Chile}

\author[0000-0003-3474-1125]{George C. Privon}
\affiliation{Department of Astronomy, University of Florida, 211 Bryant Space Science Center, Gainesville, FL 32611, USA}

\author{Neil Nagar}
\affiliation{Universidad de Concepci\'{o}n, Departamento de Astronom\'{\i}a, Casilla 160-C, Concepci\'{o}n, Chile}

\author[0000-0001-7421-2944]{Anne M. Medling}
\affiliation{Ritter Astrophysical Research Center, University of Toledo, Toledo, OH 43606, USA}

\author[0000-0002-1912-0024]{Vivian U}
\affiliation{Department of Physics and Astronomy, 4129 Frederick Reines Hall, University of California, Irvine, CA 92697, USA}

\author[0000-0002-8686-8737]{Franz E. Bauer}
\affiliation{Instituto de Astrof\'{\i}sica  and Centro de Astroingenier{\'{\i}}a, Facultad de F\'{i}sica, Pontificia Universidad Cat\'{o}lica de Chile, Casilla 306, Santiago 22, Chile}
\affiliation{Millennium Institute of Astrophysics (MAS), Nuncio Monse{\~{n}}or S{\'{o}}tero Sanz 100, Providencia, Santiago, Chile}
\affiliation{Space Science Institute, 4750 Walnut Street, Suite 205, Boulder, Colorado 80301}

\author[0000-0003-0522-6941]{Claudia Cicone}
\affiliation{INAF --- Osservatorio Astronomico di Brera, Via Brera 28, I-20121 Milano, Italy}
\affiliation{Institute of Theoretical Astrophysics, University of Oslo, Postboks 1029, 0315 Oslo, Norway}

\author[0000-0003-0057-8892]{Loreto Barcos Mu\~noz}
\affiliation{Joint ALMA Observatory, Alonso de C\'{o}rdova 3107, Vitacura, Santiago, Chile}
\affiliation{Department of Astronomy, University of Virginia, 530 McCormick Road, Charlottesville, VA 22904, USA}
\affiliation{National Radio Astronomy Observatory, 520 Edgemont Road, Charlottesville, VA 22903, USA}

\author[0000-0003-2638-1334]{Aaron S. Evans}
\affiliation{Department of Astronomy, University of Virginia, 530 McCormick Road, Charlottesville, VA 22904, USA}
\affiliation{National Radio Astronomy Observatory, 520 Edgemont Road, Charlottesville, VA 22903, USA}

\author{Francisco Muller-Sanchez}
\affiliation{Physics Department, University of Memphis, Memphis, TN 38152, USA}

\author{Julia M. Comerford}
\affiliation{Department of Astrophysical and Planetary Sciences, University of Colorado, Boulder, CO 80309, USA}

\author{Lee Armus}
\affiliation{Spitzer Science Center, California Institute of Technology, Pasadena, CA 91125, USA}

\author{Chin-Shin Chang}
\affiliation{Joint ALMA Observatory, Alonso de C\'{o}rdova 3107, Vitacura, Santiago, Chile}

\author{Michael Koss}
\affiliation{Eureka Scientific, 2452 Delmer Street, Suite 100, Oakland, CA 94602-3017, USA}

\author{Giacomo Venturi}
\affil{Instituto de Astrof{\'i}sica, Facultad de F{\'i}sica, Pontificia Universidad Cat{\'o}lica de Chile, Casilla 306, Santiago 22, Chile}

\author{Kevin Schawinski}
\affiliation{Institute for Particle Physics and Astrophysics, ETH Zurich, Wolfgang-Pauli-Str. 27, CH-8093 Zurich, Switzerland}
\affiliation{Modulos AG, Technoparkstrasse 1, CH-8005 Zurich, Switzerland}

\author[0000-0002-0930-6466]{Caitlin Casey}
\affiliation{Department of Astronomy, The University of Texas at Austin, 2515 Speedway Boulevard Stop C1400, Austin, TX 78712, USA}

\author[0000-0002-0745-9792]{C. Megan Urry}
\affiliation{Yale Center for Astronomy \& Astrophysics, Physics Department, New Haven, CT 06520, USA}

\author{David B. Sanders}
\affiliation{Institute for Astronomy, 2680 Woodlawn Drive, University of Hawaii, Honolulu, HI 96822, USA}

\author{Nicholas Scoville}
\affiliation{Cahill Center for Astrophysics, California Institute of Technology, 1216 East California Boulevard, Pasadena, CA 91125, USA}

\author{Kartik Sheth}
\affiliation{NASA Headquarters, 300 E Street SW, Washington DC 20546, USA}

\begin{abstract}
We present the highest resolution --- 15 pc (0.03'') --- ALMA $^{12}$CO(2-1)  line emission and  1.3mm continuum maps, tracers of the molecular gas and dust, respectively, in the nearby merging galaxy system NGC 6240, that hosts two supermassive 
black holes growing simultaneously. These observations provide an excellent spatial match to existing Hubble optical and near-infrared observations of this system. A significant molecular gas mass, $\sim$9$\times$10$^9$M$_\odot$, is located in between 
the two nuclei, forming a clumpy stream kinematically dominated by turbulence, rather than a smooth rotating disk as previously assumed from lower resolution data. 
Evidence for rotation is seen in the gas surrounding the southern nucleus, but not in the northern one. Dynamical 
shells can be seen, likely associated with nuclear supernovae remnants. We further detect the presence of significant high velocity outflows, some of them reaching velocities $>$500 km/s, affecting a significant fraction, $\sim$11\% of the molecular gas in the nuclear region. Inside 
the spheres of influence of the northern and southern supermassive black holes we find molecular masses of 7.4$\times$10$^8$M$_\odot$ and 3.3$\times$10$^9$M$_\odot$, respectively. We are thus directly imaging the reservoir of gas that can accrete onto each 
supermassive black hole. These new ALMA maps highlight the critical need for high resolution observations of molecular gas in order to understand the feeding of supermassive black holes and its connection to galaxy evolution in the context of a major galaxy merger.
\end{abstract}

\keywords{galaxies: interactions, active, Seyfert, indvidual (NGC 6240)}


\section{Introduction} \label{sec:intro}

In our current understanding of galaxy evolution, major galaxy-galaxy mergers play a fundamental role \citep{sanders88,hopkins08}, both for the cosmic history of star formation and the growth of their central supermassive black hole (SMBH). Computational 
simulations \citep[e.g.,][]{barnes91} show that the dynamical interactions between merging galaxies cause gas to lose angular momentum and fall into the center of each nucleus, triggering significant episodes of both star formation and 
SMBH accretion, generating a so-called luminous active galactic nucleus (AGN; \citealp{treister12,glikman15,kocevski15,fan16,trakhtenbrot17,weston17,donley18,goulding18,weigel18}). In this scenario, the relatively short-lived phase, $\sim$1-100 Myrs, in which both SMBHs are 
growing simultaneously (dual AGN), is the critical period when black hole and star formation activity are the most vigorous \citep{vanwassenhove12,blecha13,blecha18}.

NGC 6240 is often considered as the prototypical dual AGN. Based on existing optical images, NGC 6240 was classified as a late-stage compact merger  \citep[e.g.,][]{devaucouleurs64}. X-ray observations carried out using Chandra 
by \citet{komossa03} and later by \citet{wang14}  revealed two nuclei in hard X-rays, 2-8 keV, separated by $\sim$2$''$ ($\sim$950 pc), each with prominent Fe K$\alpha$ emission, clear indicator of AGN activity \citep[e.g.,][]{nandra94}. The masses of each SMBH have been 
estimated dynamically with resolved integral-field spectroscopy at $\sim$9$\times$10$^8$M$_\odot$ \citep{medling15}. While the observed E$<$10 keV X-ray luminosities for each nucleus are relatively modest, $L_X$$\sim$10$^{42}$~erg/s, the Swift/BAT observations 
at $E$=14-195 keV \citep{baumgartner13} show that their intrinsic, absorption-corrected, luminosities (and hence SMBH accretion rates) are significantly higher. NGC 6240 is also characterized by its very high IR luminosity of 0.9$\times$10$^{12}~$L$_\odot$ \citep{veilleux09}, that places
it right at the lower limit for the Ultra-Luminous IR Galaxy (ULIRG) classification. Additionally, earlier observations of the molecular gas in this merging system indicate that most of the gas can be found in between the two nuclei \citep[][and many others since]{bryant99}, and not 
around them. Hence, most of the IR emission appears to be attributed to star-formation processes rather than the AGN, likely arising from a significant starburst with an age $<$20 Myrs, as revealed by near-IR \citep{tecza00,engel10} and mid-infrared spectroscopy \citep{inami13,stierwalt13}. 
Therefore, NGC 6240 is the ideal laboratory to study the interplay between AGN and star formation in major galaxy mergers. 

As argued by \citet{tacconi99} and \citet{engel10}, NGC 6240 might represent an earlier evolutionary stage in which the gas is still in the process of settling in a central thin disk between the two nuclei. In the next stage, it will probably experience a major starburst event such as those 
observed in Arp 220 and other ULIRGs \citep{downes98}. As discussed extensively in the past and clearly confirmed in this work, NGC 6240 is a very complex system. As such, it is not surprising that the determination of even basic physical parameters in this
galaxy are still heavily debated in the literature. For example, the estimated global star formation rate ranges from a lower value of 25$M_\odot$/yr \citep{engel10} to $>$100$M_\odot$/yr \citep{howell10}, while specific nuclear regions have been claimed to have values up 
to $\sim$270$M_\odot$/yr \citep{pasquali04}.

Previous sub-arcsecond CO(2-1) observations of NGC 6240 using the IRAM Plateau de Bure interferometer carried out by \citet{tacconi99} indicated that most of the molecular gas was located between the two nuclei. Most of this gas was described to be  
in a turbulent, rotating, thick disk. This central gas concentration of $\sim$2-4$\times$10$^9$M$_\odot$ represents $\sim$30-70\% of the  dynamical mass of the system. This is however based on the assumed rotation in the central CO source. Similarly, later observations using the 
Submillimeter Array (SMA) by \citet{iono07}  found that the $^{12}$CO(3-2) emission is extended on a 4$''$ scale (corresponding to $\sim$1.9 kpc), while in contrast the HCO$^+$ (4-3) is strongly concentrated in the central kpc between the two nuclei. Large velocity offsets of up 
to $\sim$800 km/s were already observed in the $^{12}$CO(1-0) emission of NGC 6240, thanks to IRAM observations at $\sim$1$''$ resolution \citep{feruglio13}. This provides a clear indication that AGN and/or supernova feedback is taking place in this system. Indeed, 
recently \citet{mueller-sanchez18} reported the presence of kpc-scale ionized outflows in NGC 6240. Non-spatially overlapping ionized outflow rates of $\sim$10M$_\odot$/yr and 75M$_\odot$/year were attributed to star formation and AGN feedback respectively. Combined, they 
are comparable to the measured star formation rate in the system \citep{mueller-sanchez18}.  

From HI observations, the recession velocity of the North and South nuclei are 7440 km/s and 7258 km/s, respectively \citep{beswick01}. \citet{baan07} posit that it is most plausible that the south nucleus lies behind the northern one with the line joining the nuclei at 13\arcdeg\ to the 
line of sight, i.e. a projection factor of 4.3. The projected separation between nuclei ($\sim$1.51\arcsec\ at PA $\sim$19\arcdeg) would thus correspond to a true distance of $\sim$3.2~kpc. The stellar velocity field of the Southern galaxy suggests a NW-SE rotation (PA = 130\arcdeg to
the receding side; inclination $\sim$60\arcdeg) and the Northern galaxy suggests a southwest-northeast rotation (PA = 61\arcdeg to the receding side; inclination i = 33\arcsec\ to 52\arcdeg) \citep{baan07,medling11,medling14}. Shells and bubbles observed at radio wavelengths in 
this merger are argued to originate from outflows driven by the South nucleus \citep{heckman90,baan07}. Centimeter VLBI observations of NGC~6240 \citep{hagiwara11} have not only detected the two nuclei (previous nomenclature used: N1 for the South nucleus and N2 for the North 
nucleus) but also two radio supernova: RS1 which is 35~mas southwest of the South nucleus and RS2 which is 0\farcs287 northeast (PA 28\arcdeg) of the south nucleus. 

NGC 6240 was the target of several Atacama Large Millimiter/submillimiter Array (ALMA) observations. \citet{scoville15} used dust continuum observations at 350 GHz (band 7) and 696 GHz (band 9) to derive a nuclear interstellar medium (ISM) mass of 1.6$\times$10$^9$M$_\odot$ in a
$\sim$200pc radius. More recently, \citet{saito18} presented ALMA CO(2-1) band 6 observations of this system at moderate resolutions, $\sim$1$\times$0.5 arcsec$^2$,  finding --- in addition to the central concentration --- extended high-velocity, $\sim$2,000 km/s, components, with
a total mass of 5$\times$10$^8$M$_\odot$. Similarly, four broad components located $\sim$1-2 kpc away from the center of the system were reported. Finally, \citet{cicone18} combined [CI](1-0) and CO(2-1) ALMA observations with IRAM Plateau de Bure Interferometer CO(1-0) maps
to trace molecular outflows in the central 6$\times$3 kpc$^2$ region of NGC 6240. They found outflowing molecular gas peaking between the two SMBHs, and extending by several kpc along the east-west direction. The total H$_2$ outflow rate is 2500$\pm$1200 M$_\odot$/yr, where the error 
includes the uncertainty on the $\alpha_{\textnormal{CO}}$ factor, hence confirming that a combination of AGN and star formation feedback is required to drive the observed outflows. Similar to the values of the star formation rate, the mass outflow rate in different parts of the system is very 
relevant for this work and can vary by more than an order of magnitude, ranging from $\sim$230$M_\odot$/yr \citep{saito18} up to $\sim$2500$M_\odot$/yr in the central region \citep{cicone18}, due to different assumptions for the geometry of the outflow and the position where the outflow
is measured. The range in these measurements will be considered in our analysis and interpretation of this system.

Here, we present new ALMA observations of the $^{12}$CO(2-1) rotational transition with the highest angular resolution to date, using up to 15 km baselines. Section \S2 presents the technical properties of the data and details of the data processing and reduction. Section \S3 shows
the results including emission, kinematic and velocity dispersion maps. These results are discussed in section \S4, while the conclusions are presented in \S5. Throughout this paper, we assume a $\Lambda$CDM cosmology  with $h_0$=0.7, 
$\Omega_m$=0.27 and $\Omega_\Lambda$=0.73 \citep{hinshaw09}.

\section{ALMA Data} \label{sec:data}

ALMA observed NGC 6240 as part of its cycle 3 program 2015.1.00370.S (PI: E. Treister). The $^{12}$CO(2-1) at rest-frame 230.538 GHz was targeted as a tracer of the molecular gas in the central region of the system. The achieved angular resolution is $\sim$0.03$''$, corresponding 
to a physical scale of $\sim$15 pc, $\sim$10$\times$ higher than previous observations of this and similar molecular transitions \citep{tacconi99,iono07,wilson08,scoville15,saito18,cicone18}. Furthermore, this is comparable and slightly better than existing HST optical imaging \citep{gerssen04} 
and $>$2$\times$ sharper than near-IR HST maps \citep{scoville00}.  Previous ground-based observations assisted by adaptive optics, presented by \citet{max07}, reached similar spatial resolutions at near-IR wavelengths. The main technical goal of our program was to obtain the highest 
possible resolution maps of the $^{12}$CO(2-1) emission, while at the same time recovering most of the total flux at large scales by combining three different array configurations (baselines from 15 meters to 15 km). The observations were divided on two science goals, comprised of three different 
configuration groups overall. Data from program 2015.1.00003.S \citep{saito18} were also included to improve the recovery of the large scale structure from intermediate-length baselines.

\subsection{Description of Observations}

Table~\ref{tab:obsspecs} details the properties of each observation considered in this study. Despite not being used in the final imaging for reasons listed below, the long-baseline execution block (EB) Xac5575/X8a5f is also listed. All the observations described, except Xb4da9a/X69a, share 
the same spectral setup. The total ALMA bandwidth of 7.5\,GHz was divided into four spectral windows (SPWs) of 1875 MHz each and a spectral resolution of 1.953 MHz. In all observations, one of these SPWs was specifically tuned to study the redshifted $^{12}$CO(2-1) transition, 
centered at a frequency of 225.1\,GHz. The remaining ones were used to trace the mm continuum centered at frequencies of 227.5\,GHz, 239.9\,GHz and 241.9\,GHz, with the second one partly covering the CS\,(5-4) line. The Xb4da9a/X69a EB 
covered the $^{12}$CO(2-1) and CS\,(5-4) transitions with two SPWs centered at 224.9 and 239.8\,GHz with spectral resolutions of 1.953 and 3.906\,MHz, respectively. The remaining two SPWs traced the continuum at 221.9 and 237.9\,GHz with spectral resolutions of 7.813\,MHz.

\begin{sidewaystable}
	\centering
	\begin{tabular}{| p{28mm} c p{13mm}p{22mm} rccccr|} \hline
		date & EB & Basel$_{min}^{max}$ & res/MRS & N$_{\rm ANT}$ & Flux & Bandpass & Phase & Check & ToS \\\relax
		[UTC] & [uid://A002/...] & [m] & [arcsec] &  &  &  &  & & [min] \\
		\hline
		2015-11-01 \newline 17:24---18:38UT & Xac5575/X8a5f & 14969 \newline 84.7 & 0.024 \newline 0.38 & 41 & J1550+0527 & J1550+0527 & J1651+0129 & J1649+0412 & 32 \\
		\hline
		2016-01-29 \newline 10:46---11:02UT & Xaf985b/Xf3f & 331 \newline 15.1 & 0.97 \newline 9.3 & 44 & J1550+0527 & J1550+0527 & J1651+0129 & J1659+0213 & 1.1 \\
		\hline
		2016-05-31 \newline 04:09---04:33UT & Xb3c4ab/Xbef & 784 \newline 15.1 & 0.44 \newline 5.12 & 39 & J1550+0527 & J1751+0939 & J1651+0129 & J1659+0213 & 2.1 \\
		\hline
		2016-07-28 \newline 04:17---04:41UT & Xb5fdce/X79d & 1124 \newline 15.1 & 0.26 \newline 3.51 & 42 & Pallas & J1751+0939 & J1651+0129 & J1659+0213 & 2.1 \\
		\hline
		2017-09-29 \newline 23:39---00:49UT & Xc5148b/X127e & 14969 \newline 41.4 & 0.023 \newline 0.46 & 43 & J1751+0939 & J1751+0939 & J1651+0129 & J1649+0412 & 35 \\
		\hline
		\hline
		2016-06-26 \newline 13:29---13:53UT & Xb4da9a/X69a & 784 \newline 15.1 & 0.43 \newline 5.41 & 42 & Titan & J1550+0527 & J1651+0129 & \ldots & 4.2 \\
		\hline
	\end{tabular}
	\caption{The characteristics of each EB considered in this study: the date and time of observation; the EB UID code; the maximum and minimum baseline length; the expected spatial resolution and maximum recoverable scale (res/MRS); number of antennas; the sources used for flux, bandpass, and phase calibrator together with the check source; and the time on source (ToS), i.e., NGC 6240. Note that uid://A002/Xb4da9a/X69a was observed as part of 2015.1.00003.S \citep{saito18}.}
	\label{tab:obsspecs}
\end{sidewaystable}

\subsection{Data Analysis}

For the data analysis, the Common Astronomy Software Applications (CASA, v5.4; \citealp{mcmullin07}) was used. In general, the standard ALMA data reduction\footnote{Available at \url{https://almascience.nrao.edu/processing/science-pipeline}} approach was pursued,
with   some specific differences. More details about the array configuration properties, observations, processing method and imaging parameters are 
reported in appendix A.

\section{Results}\label{sec:results}

\subsection{NGC 6240 $^{12}$CO(2-1) Emission}

\begin{figure}
\plottwo{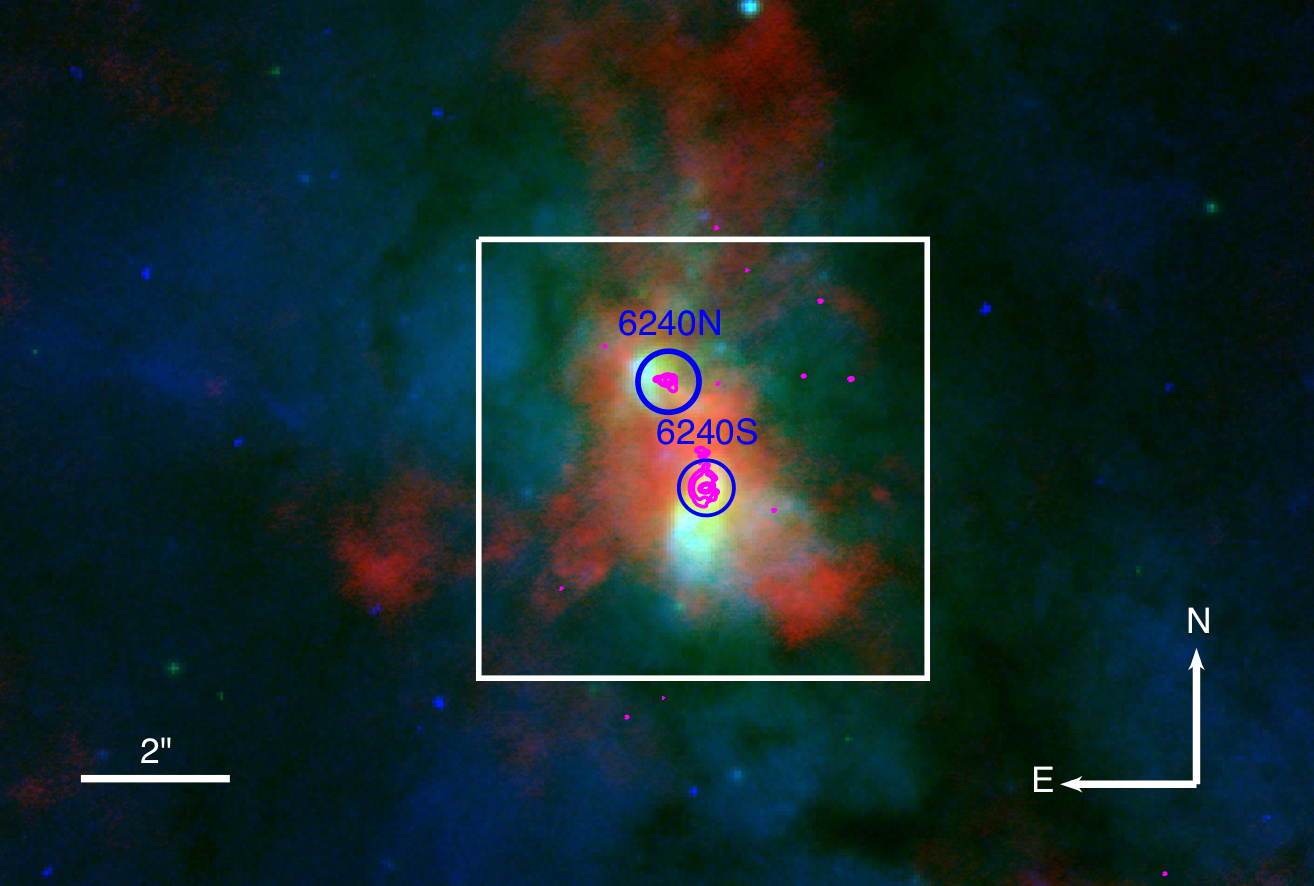}{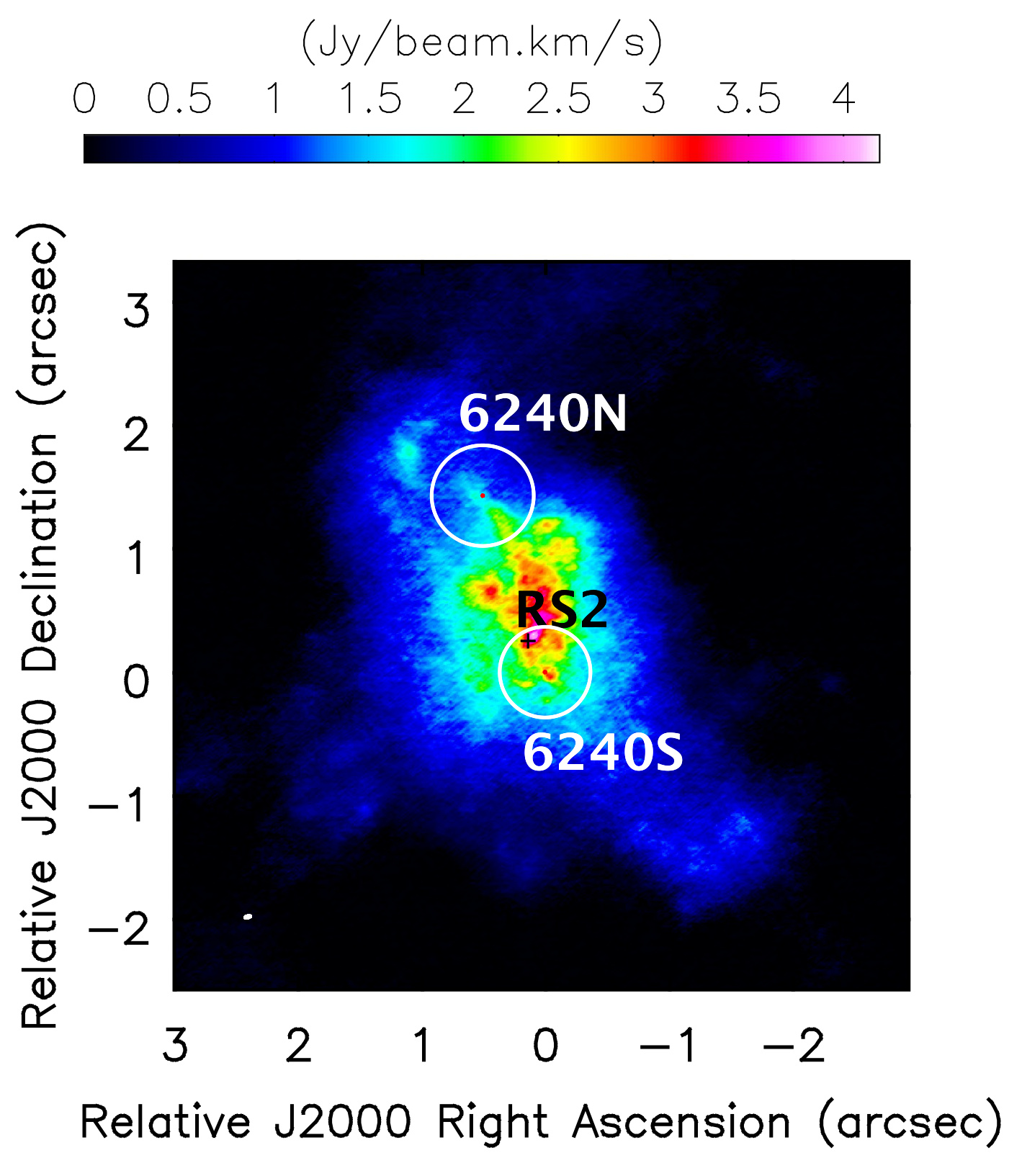}
\caption{ {\it Left panel:} Central region of the NGC 6240 system covering approximately 13$\times$8 kpc$^2$ (2$''$$\simeq$1 kpc). Blue and green colors denote images from the HST/WFPC2 camera on the F450W and F814W bands, respectively. The red color 
denotes the moment 0 ALMA $^{12}$CO(2-1) observations presented here. The blue circles mark the locations
of the North and South nuclei based on VLBI radio observations \citep{hagiwara11}. Magenta contours show the continuum emission at 235 GHz, as obtained from our ALMA data and presented in Figure~\ref{ngc6240_alma_cont}, in five logarithmic levels 
starting from 4$\times$10$^{-5}$ Jy/beam and ending at 4$\times$10$^{-3}$ Jy/beam. The white square presents the zoom region displayed on the right panel. There, we show the integrated intensity map of the
$^{12}$CO(2-1) emission, as observed by ALMA. White circles indicate the spheres of influence of each SMBH, as estimated by \citet{medling15} based on near-IR
integral field spectroscopic observations obtained with the Keck telescopes, while their location is shown by red dots. The black cross marks the position of the supernova RS2, as reported from radio observations by \citet{hagiwara11}.\label{ngc6240_alma_hst}}
\end{figure}

Figure~\ref{ngc6240_alma_hst} shows the $^{12}$CO(2-1) emission together with the optical imaging provided by the Hubble Space Telescope (HST) using the Wide Field Planetary Camera 2 (WFPC2) instrument. As described above, by combining low and high resolution observations 
we are able to recover all the emission out to scales of 30$''$. Indeed, the total measured flux is 1163$\pm$28~Jy km/s in the ALMA field of view of 25$''$ diameter, which accounts for 95\% of the total CO(2-1) when compared to the \citet{tacconi99} measurements. We can further
compare this total flux to single-dish observations. An unresolved, $^{12}$CO(2-1) flux of 1492$\pm$253~Jy km/s was reported by \citet{greve09}, implying that our ALMA observations recover 67-94\% of the single-dish flux when considering the relatively large error bars of the
\citet{greve09} measurements. As can be seen in Figure~\ref{ngc6240_alma_hst}, a significant concentration of the molecular gas is found in the central regions, consistent with findings based on previous observations \citep[e.g.,][]{tacconi99,engel10}. However, extended 
molecular emission can be detected up to $\sim$10$''$ ($\sim$5 kpc) away from the nuclear region. The bulk of the molecular gas emission in the nuclear region of the system appears to be directly connected to the material found in between the nuclei. 

This molecular CO emission does not spatially coincide with the sub-mm continuum observed by ALMA at observed-frame 235 GHz, as shown by the magenta contours  on the left panel of Figure~\ref{ngc6240_alma_hst} and in more detail in Figure~\ref{ngc6240_alma_cont}. Similarly, the 
bulk of the  $^{12}$CO(2-1) emission does not appear to overlap  with the stellar light traced by the optical and near-IR emission either, which is mostly centered around each nucleus \citep{max05,engel10}. This could be naturally explained by the effects of dust extinction in the
nuclear regions. Indeed, \citet{max05,max07,mueller-sanchez18} reported the presence of significant dust lanes in the center of the NGC 6240 system. For our work, the nuclear positions were obtained from the
Very Long Baseline Interferometry (VLBI) observations of this source, as reported by \citet{hagiwara11}. Specifically, the nuclear positions used are 16$^h$52$^m$58.924, 02$^o$24$'$04.776$''$ and 16$^h$52$^m$58.890, 02$^o$24$'$03.350$''$ for the northern and southern ones
respectively. Typical errors in these positions are 0.6 mas. Only a relatively small fraction of the molecular gas, compared to the total amount in the central region of the system, is found inside the sphere of influence of each SMBH, which was computed by \citet{medling15} to have 
radii of 235$\pm$8 pc and 212$\pm$9 pc for the northern and southern nuclei, respectively,  and are hence fully resolved in the ALMA cube. This can be used to improve existing measurements of the SMBH mass in each nucleus, by separating the SMBH and molecular gas contributions to the 
total enclosed nuclear mass, which in turn indicates that these are closer to the usual M-$\sigma$ correlation than previously indicated \citep{medling19}. Recently, \citet{kollatschny19} claimed the detection of a third nucleus in this 
system from optical integral field spectroscopy observations, close to the southern nucleus at a position of 16$^h$52$^m$58.901, 02$^o$24$'$02.88$''$. We do not detect any significant $^{12}$CO(2-1) emission at this position or its surroundings, in stark contrast with the two previously 
established nuclei.

In contrast to previous sub-mm observations at lower spatial resolutions \citep{tacconi99,saito18}, the molecular gas in the nuclear region does not appear to follow a smooth distribution. Instead, substantial clumpiness can be found on the stream that connects both nuclei, as can be 
seen in Fig.~\ref{ngc6240_alma_hst}. Similar structures in the spatial distribution of the molecular emission, in this case from $^{12}$CO(3-2) observations, have been previously presented by \citet{u11} and \citet{wang14}. These clumpy regions are natural sites where vigorous 
star formation should take place. However, according to high resolution, $\sim$0.1$''$, near-IR observations at 2.2$\mu$m obtained with the VLT/SINFONI integral field spectrograph, reported by \citet{engel10}, the bulk of the nuclear stellar mass appears to precede 
the merger and is spatially concentrated around each nucleus, certainly much more than the molecular gas traced by the ALMA observations.

In order to take advantage of the superb spatial resolution of the ALMA observations, the focus of this work is mostly on the relatively small-scale, sub-arcsec, 
structures. That being said, we can place these in context of more extended structures previously reported from CO observations. Specifically, \citet{feruglio13b} reported the detection
of CO(1-0) emission extending up to 20$''$ ($\sim$10 kpc) from the central region of the system, both in the eastward and south-westward directions. The eastward
extended structure is also detected in our ALMA data, at velocities of $\sim$-300 to -400 km/s and reaching up to $\sim$12$''$ at the edge of the field of view
of our data cube; as such we cannot verify a larger extension. We also detect the south-westward structure at a velocity of $\sim$-100 km/s. The extension
of this structure is $\sim$13$''$ from the southern nucleus, where it reaches the edge of the field of view, as in the previous case. Hence, we confirm the
detection of extended CO emission previously reported by \citet{feruglio13b}. We can further identify an additional extended structure reaching up to $\sim$7$''$
to the north-west of the northern nucleus at velocities of $\sim$100 to 200 km/s. A detailed characterization of these large scale structures is beyond the scope of this article.

\subsection{Continuum emission}
\label{sec:continuum}

\begin{figure}
\begin{center}
\plotone{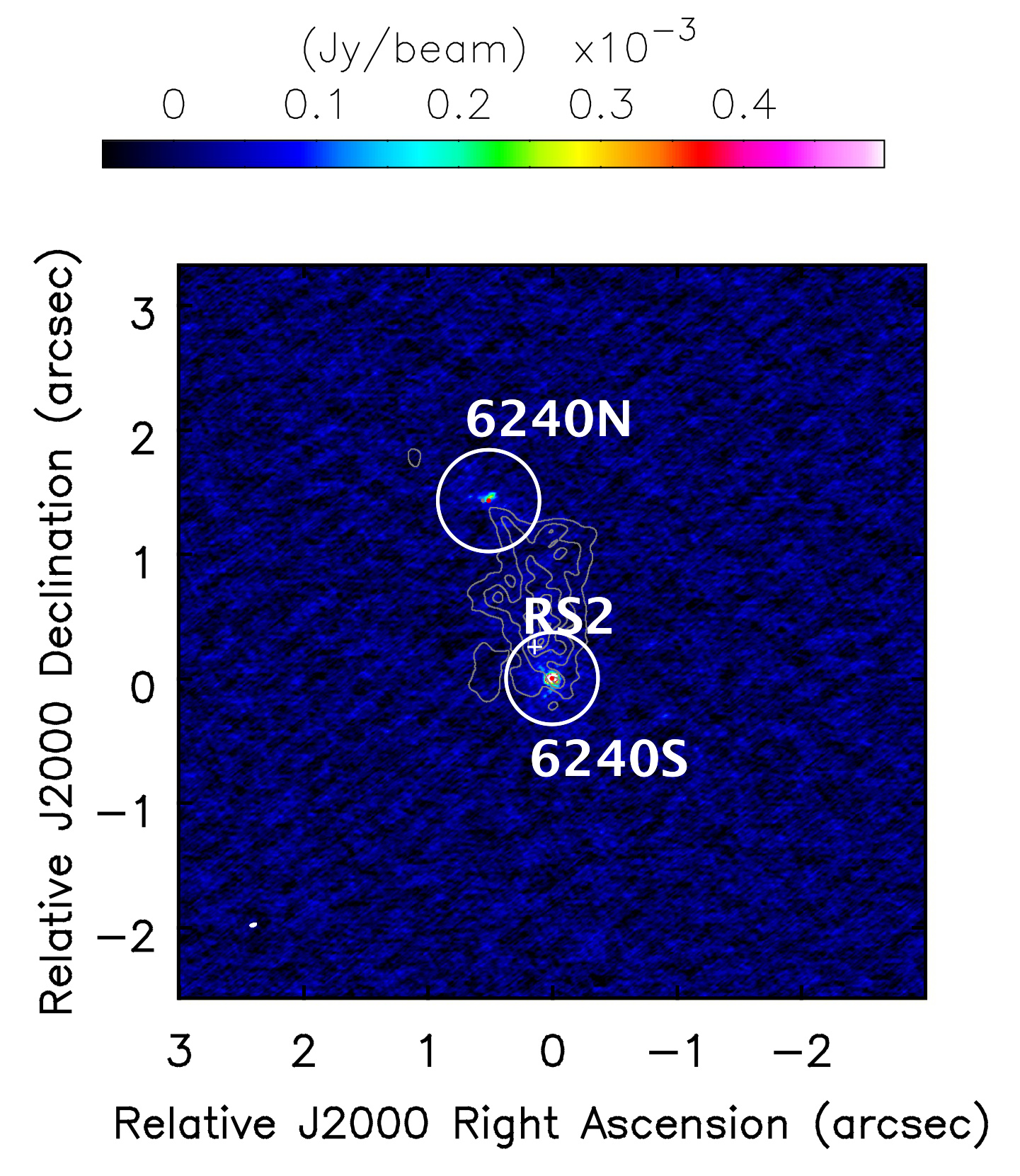}
\end{center}
\caption{Millimeter continuum emission (colorscale) in the central region of the NGC 6240 at an effective observed-frame frequency of 235 GHz, as obtained from the ALMA observations described here. The white circles have radii of 235 and 212 pc for the north and south nuclei, respectively,
corresponding to the spheres of influence of each SMBH, while the gray contours present the integrated (moment 0) $^{12}$CO(2-1) emission. The position of
RS2 is presented by the white cross. With an observed spectral index of -0.91 \citep{hagiwara11}, the expected flux of RS2 ($\sim$12 $\mu$Jy) falls well below the sensitivity 
of these ALMA observations. Each contour shows an increase of 20\% in flux starting from a base level of 0.2 Jy/beam km/s. The beam size and shape of the continuum observations 
can be seen in white in the bottom left. As can be seen, the mm continuum emission is concentrated around each SMBH and shows little overlap with the molecular gas traced by the $^{12}$CO(2-1) transition.\label{ngc6240_alma_cont}}
\end{figure}

As described in section~\ref{sec:data}, some of the ALMA spectral windows were used to study the continuum emission at $\sim$235\,GHz, corresponding to $\sim$1.25mm, which we show in Figure~\ref{ngc6240_alma_cont}.  
As can be seen there, and contrary to what was observed for the $^{12}$CO(2-1) transition that mostly traces the molecular gas, the millimeter continuum emission arising from dust is mostly concentrated around each 
nucleus. We measure continuum flux densities at an observed-frame frequency of 235 GHz of 9.92$\pm$0.16 mJy and 3.50$\pm$0.037 mJy inside 
the corresponding SMBH spheres of influence for the southern and northern nucleus, respectively. Considering now a much larger aperture with a diameter of 22$''$ 
we measure a continuum flux of 13.0$\pm$0.6 mJy. Hence, we can conclude that in this source most of the total mm continuum emission detected in our ALMA data is 
directly associated with the two nuclear regions. Furthermore, as can be seen in Figure~\ref{ngc6240_alma_cont}, the emission is strongly concentrated and 
marginally resolved. Similarly to the non-detection of CO(2-1) emission discussed in the previous section, at the flux limit of our observations we do not detect 
any sub-mm continuum emission at the location of the third nucleus reported by \citet{kollatschny19}.

The continuum fluxes reported here are significantly higher than those reported by \citet{tacconi99} at 228 GHz, by 2.1 times for the southern nucleus and 
3.5 times for the northern one, using similar apertures, $\sim$0.8$''$ in diameter. This discrepancy can likely be explained by a sparse \textit{uv} coverage by our long baseline observations
combined with their poor phase stability, which particularly affect flux measurements on relatively small scales, $\sim$1$''$, after integrating over a relatively large wavelength range. Indeed, when continuum
fluxes are measured using only the compact and intermediate array configurations described in \S\ref{compact_conf}, we obtain values of 5.19$\pm$0.06 mJy and
1.73$\pm$0.03 mJy for the southern and northern nuclei respectively, roughly consistent with the \citet{tacconi99} values. The discrepancy with the maps containing the long baselines is much smaller both if we use much smaller apertures, $<$0.5$''$, as was done by \citet{medling19} or
much larger ones. Indeed, on a 22$''$ diameter aperture we measure a continuum flux of 10.1$\pm$0.13 mJy, again showing that most of the 
mm continuum emission in this system is associated with the two nuclei. Therefore, in what follows, we only use these latter fluxes, as clearly better \textit{uv} coverage and phase stability are required to measure continuum fluxes at relatively small 
scales. The continuum emission map presented in Figure~\ref{ngc6240_alma_cont} is however not significantly affected, as this ``extra'' flux has a relatively low surface density. Similarly, by comparing with
previously reported $^{12}$CO(2-1) fluxes at a range of scales and resolutions we concluded that our line measurements have not been significantly affected. Hence, this issue is only relevant for continuum studies that require one to integrate flux densities 
over a large wavelength range. Note that the use of feathering on a per-channel basis, when retrieving the $^{12}$CO(2-1) emission map, prevented this low-surface density to be parsed to the final combined map.

Previously, \citet{nakanishi05} presented continuum observations at frequencies of 87 and 108 GHz, with synthesized beam sizes 
of $\sim$2$''$ and $\sim$4$''$ and total measured fluxes of 16.6 and 10.8 mJy at each frequency, respectively. Even at lower spatial resolution, 
both \citet{nakanishi05} and \citet{tacconi99} were able to identify an offset between the continuum and the line emission, which is further corroborated by our ALMA data. 
The spectral shape in these data is given by a power law with index $\alpha$=-0.81, consistent with synchrotron emission from non-thermal sources such as 
AGN or supernova remnants. Our ALMA observations reveal that most of the continuum emission arises from the surroundings of each SMBH and is significantly brighter than the extrapolation of the
spectrum at low frequencies, as presented in section 4.2, thus  strongly suggesting that this radiation dominated by dust heated by AGN activity.

NGC~6240 has an existing $880~\mu m$ single-dish dust continuum measurement of $F_{880}=133 \pm 40$ mJy \citep{wilson08}. Adopting a dust emissivity index 
$\beta=1.8$ \citep{planck11} this predicts a flux density at 235 GHz of $32\pm10$ mJy. Hence our observations recover approximately 41\% of the expected continuum 
flux. If we assume this missing flux is concentrated in a central 2\arcsec~diameter aperture, the predicted flux per beam is $5~\mu$Jy, below our detection threshold. 
The missing flux is hence consistent with being due to insufficient surface brightness sensitivity in these continuum observations.

\subsection{Gas kinematics}

\begin{figure}
\plotone{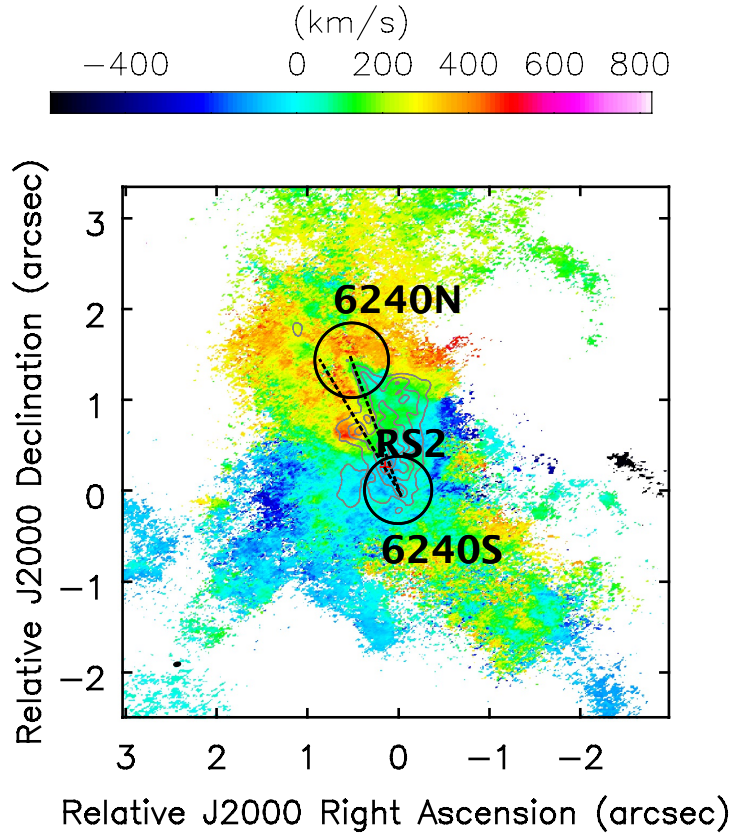}
\vspace{-7.5cm}
\caption{Velocity (moment 1) map for the $^{12}$CO(2-1) emission, from the ALMA data presented here. Black circles show $\sim$200 pc diameter circles, corresponding to the spheres of influence for each SMBH, as described in Figure~\ref{ngc6240_alma_hst}. The red cross marks the position of RS2, while
the two dashed lines show the locations connecting the two nuclei and between the southern nucleus and RS2 where position-velocity diagrams were extracted,
as shown in Figures~\ref{alma_nuclear_pv} and \ref{alma_nuclear_pv028} respectively. A hint of a velocity gradient, ranging from $\sim$300 km/s in the northern 
region to $\sim$-150 km/s can be seen in the stream connecting both nuclei. In addition, we can see high velocity regions, at $>$500 km/s, as those described 
in Figure~\ref{alma_outflow}, corresponding to a molecular gas outflow.\label{alma_velmap}}
\end{figure}

\begin{figure}
\begin{interactive}{animation}{ngc6240_apj_v2.mp4}
\plotone{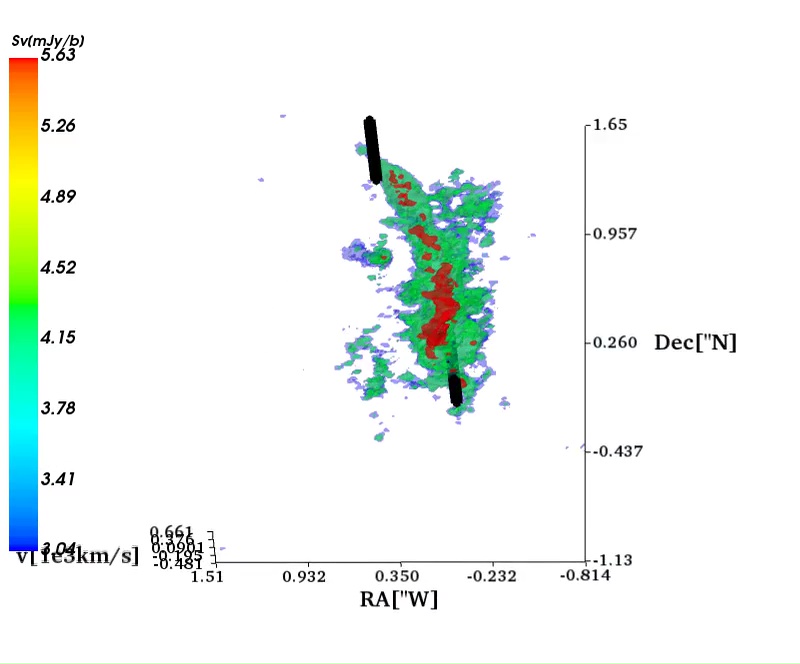}
\end{interactive}
\caption{Still frame from the exploration of the ALMA data cube available as an animation in the electronic version of the journal. Three dimensional axes show the spatial coordinates, RA and Dec relative to the center of the cube, and velocity in the line of sight. The
{\it black lines} at all velocities show the positions of the north and south nuclei. The decoupled structure at high velocity, described in section 4.3, can be seen in the animation, along with the bulk of the molecular gas in the central region of the system.\label{co21_animation}}
\end{figure}

The ALMA observations also yield information about the kinematics of the molecular gas. Figure~\ref{alma_velmap} shows the velocity map for the central region of NGC 6240 as traced by the $^{12}$CO(2-1) emission. In contrast to previous
claims \citep{tacconi99,bryant99,engel10}, there is no clear evidence for a rotating disk in the inter-nuclear region. In fact, no well-defined structures are obviously visible in the moment 1 map. A similar conclusion, albeit using lower resolution data,
was reached before by \citet{cicone18}. In this work, they reported that the typical butterfly pattern, a smoking gun for a rotating disk, could not be seen; it is not present in our data either. Instead, they concluded that the kinematics in the nuclear region were
dominated by outflows. The higher resolution ALMA data appears consistent with a significant high velocity outflow, described in more detail in section \ref{sec:outflow}, visible directly south of the northern nucleus. A tentative velocity gradient can also be observed between 
the two nuclei. Indeed, such a velocity gradient, of $\sim$100 km/s, was previously reported based on sub-arcsecond $^{12}$CO(3-2) observations by \citet{iono07} and \citet{u11}. This gradient was also found in the H$_2$ observations of this region reported by \citet{mueller-sanchez18}, who 
interpreted it as a perturbed turbulent rotational disk. To facilitate the visualization of the overall kinematical structure of the cold molecular gas in the nuclear region of NGC 6240, Figure~\ref{co21_animation} presents a snapshot of an animation available on the electronic version of
this article, exploring the three dimensional spatial and velocity cube traced by our ALMA observations.

\begin{figure}
\plotone{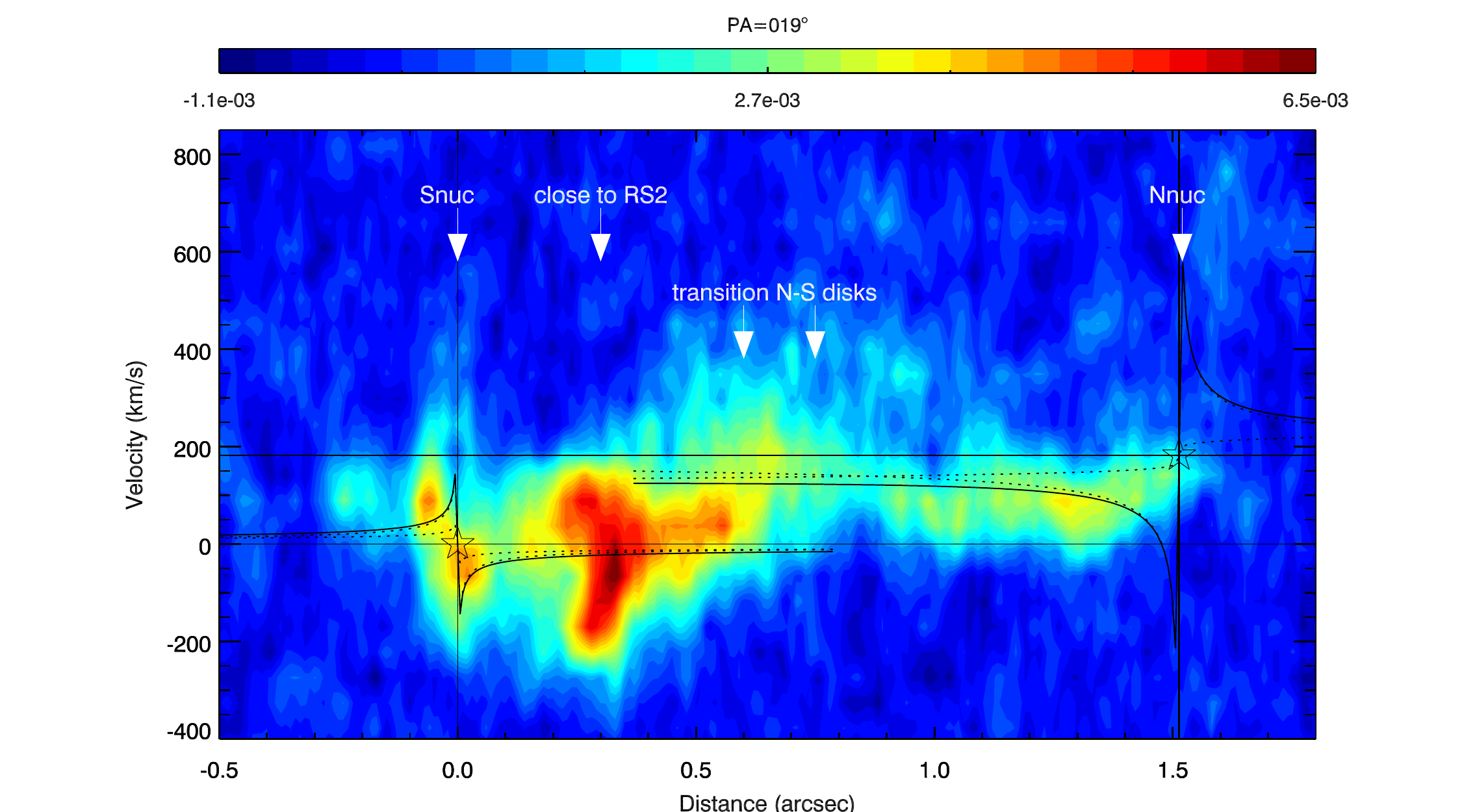}
\caption{Position-velocity diagram for the $^{12}$CO(2-1) emission along a PA of 19\arcdeg, centered on the southern nucleus and passing through the northern one. The southern 
(at zero position and velocity) and northern (offset $\sim$1.5$''$ in position and $\sim$182 km/s in velocity) nuclei are marked by open stars. 
To guide the eye, vertical and horizontal lines mark the positions and recessional velocities, respectively, of the two nuclei. Also to guide the eye, we show for each nucleus the predictions of the velocity fields for the
galaxy-only, black hole only ({\it dotted curves}), and total (galaxy plus black hole) enclosed mass ({\it solid curves}; see text).\label{alma_nuclear_pv}}
\end{figure}

In order to better understand the kinematic structure of the molecular gas stream connecting the two nuclei, Figure~\ref{alma_nuclear_pv} presents a position-velocity (p-v) diagram along a slit connecting the two nuclei. For illustration, we overplot the velocity field predictions of the respective 
stellar-disk only and stellar-disk plus black hole in the Northern and Southern nuclei, using parameters from \citet{medling11} and \citet{medling15}, and assuming for simplicity that the PA and inclination of the inner gas disks around the north and south supermassive black holes follow the same 
kinematics as the larger scale stellar disk\footnote{This need not be the case, but is assumed for simplicity since there is insufficient information to determine the kinematics of the gas from the CO(2-1) emission alone.}. First, near the southern nucleus, located inside the sphere of influence of the southern SMBH we can see a strongly localized gradient over $\sim$0.2$''$ spanning a range of approximately $\pm$200 km/s. The CO velocities here do not follow the rotating stellar disk model posited by \citet{medling15}. This gas thus either follows a different orientation or projection, or traces outflowing nuclear gas (in which case the near side of the Southern stellar disk (S) is to the northeast). Interpreting the northeast side of the Southern galaxy as the near side would be consistent with the
model of \citet{baan07} in which the Southern stellar disk is behind the Northern one (N). Extrapolating this further, one could posit that the far side of the N stellar disk is to the southeast, as this overall orientation would most easily explain the observed internuclear bridge of gas. 
At offsets of 0.1$''$ to 0.6$''$ north-east of the southern nucleus, the CO roughly follows the rotating stellar disk model of \citet{medling11} except for the high dispersion region at offset $\sim$0.3$''$ north of the S nucleus. This structure is by far the most 
prominent feature seen in the p-v diagram, as it includes the highest flux peak for the inter-nuclear $^{12}$CO(2-1) emission and extends over 400 km/s in velocity space. It is strongly localized spatially, spanning roughly $\sim$0.2$''$. This high dispersion region is located at an offset
where the slit passes very close to the radio supernova RS2 (which likely traces a starbursting region), detected at mas-scale resolution in the radio by \citet{hagiwara11}, and its peculiar molecular kinematics are discussed further below.  At offsets between 0.9$''$ to 1.5$''$ north of the S
nucleus, the gas roughly follows the predictions of the rotating Northern stellar disk of \citet{medling15}, though there are extra wiggles and a lack of Keplerian rotation inside the posited sphere of influence of the N nucleus black hole. Finally, at offsets between 0.5$''$ to 0.8$''$ north of the S nucleus
we see a clear transition region as the gas connects from one of the claimed disks to the other. In summary, the gas bridge between the nuclei appears to connect them via the near side of the (behind) Southern stellar disk and the far side of the (in front) Northern one, and while some of 
the gas may be roughly consistent with the expected dynamics of the two nuclear cores disks, there are clear streaming kinematics in the arm plus outflow signatures (in the S nucleus and near RS2) and a lack of Keplerian rotation near the N nucleus.

\begin{figure}
\begin{center}
\includegraphics[scale=1.0]{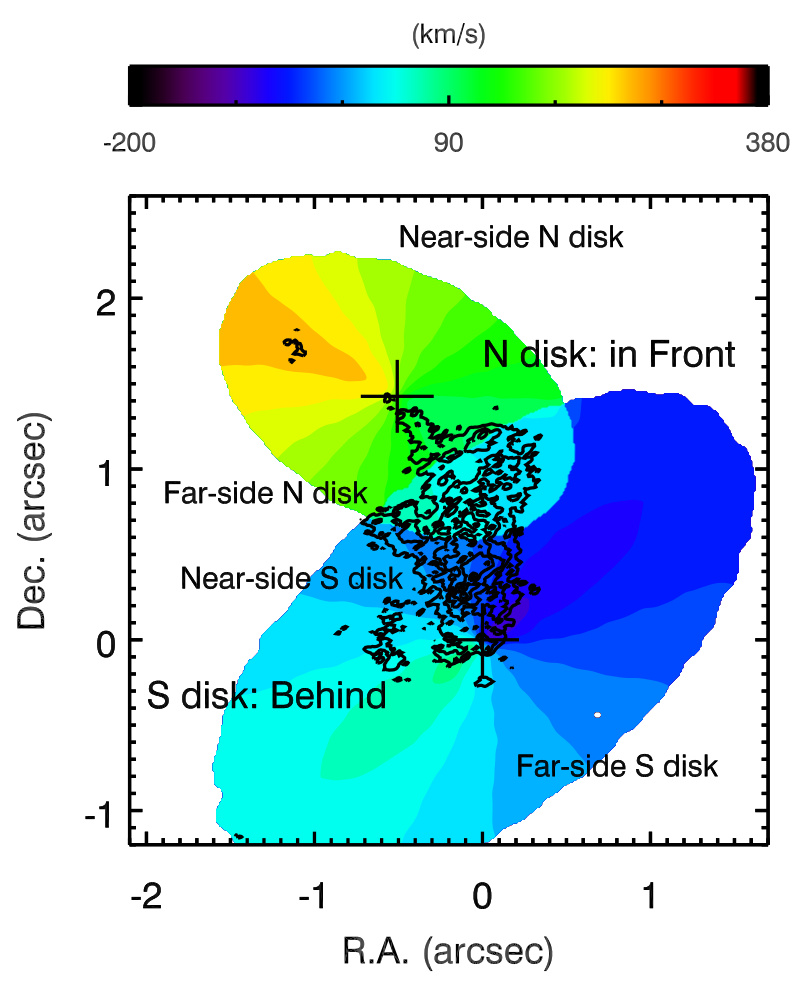}
\end{center}
\vspace{-0.5cm}
\caption{Illustration of the morphology and velocity (color; following the color bar, in km/s, at the top; the color scale is binned at 10km/s) of the inner stellar disks of the North
(N) and South (S) components in NGC~6240, with the integrated $^{12}$CO(2-1) emission overlaid in black contours.  Nuclear positions
are marked with crosses. We follow the model of \citet{baan07}, in which the N disk is closer to us and the S disk further away. There appear to be a relatively 
continuous velocity structure traced by CO as the gas kinematics transitions from the S stellar disk to that of the N stellar disk. Figure~\ref{alma_nuclear_pv} suggests that the two disks approach each other, and perhaps interact. Based on this, we mark the near- and 
far-side of each stellar disk. The N and S stellar disks shown are (for illustration) 2.2$''$ ($\sim$1.1~kpc) and 3.8$''$ (1.9~kpc) in diameter. When the disks overlap (for our line of sight; note that this does
not necessarily mean that they intersect, as discussed in the text) we show the average of the two disk velocities. The CO morphology appears to 
follow the line of overlap of the two disks, supporting a scenario in which they truly intersect. The best-fit stellar density profiles of \citet{medling15} are significantly 
different in the N and S stellar disk, with the former having a shallower ($\rho$$\propto$r$^{-0.63}$) slope than the latter ($\rho$$\propto$r$^{-1.5}$). This results in 
significantly different shapes for the  stellar-disk-only rotation curves in the two nuclei.\label{model_mom1}}
\end{figure}

Fig.~\ref{model_mom1} illustrates the projected geometry of the inner stellar disks and the CO emission, following the interpretation from \citet{baan07}, based on HI absorption, placing the N disk in front of the S one. In this case, the relatively continuous velocity
structure of the gas as it transitions between the two stellar disks (Fig.~\ref{alma_nuclear_pv}) suggests that the disks intersect each other, i.e. the NE of the S disk is the near side, and the SE of the N disk is the far side. This is what we present in Figure~\ref{model_mom1}.
In this case, both the N and S disk are rotating counter-clockwise on the sky. The E-W extension in CO in the area between the two nuclei likely traces the area where both disks actually intersect (rather than just overlap to our sight-line). In the \citet{baan07} model, the radial 
distance between the two nuclei is $\sim$3.3~kpc, so that the stellar disks illustrated in the figure would overlap by line-of-sight projection but not actually physically intersect. If this radial distance was decreased to $<$2.4~kpc, then the two illustrated stellar disks would actually intersect over the ridgeline 
at which we see the maximum E-W extension of the CO emission.

\begin{figure}
\plotone{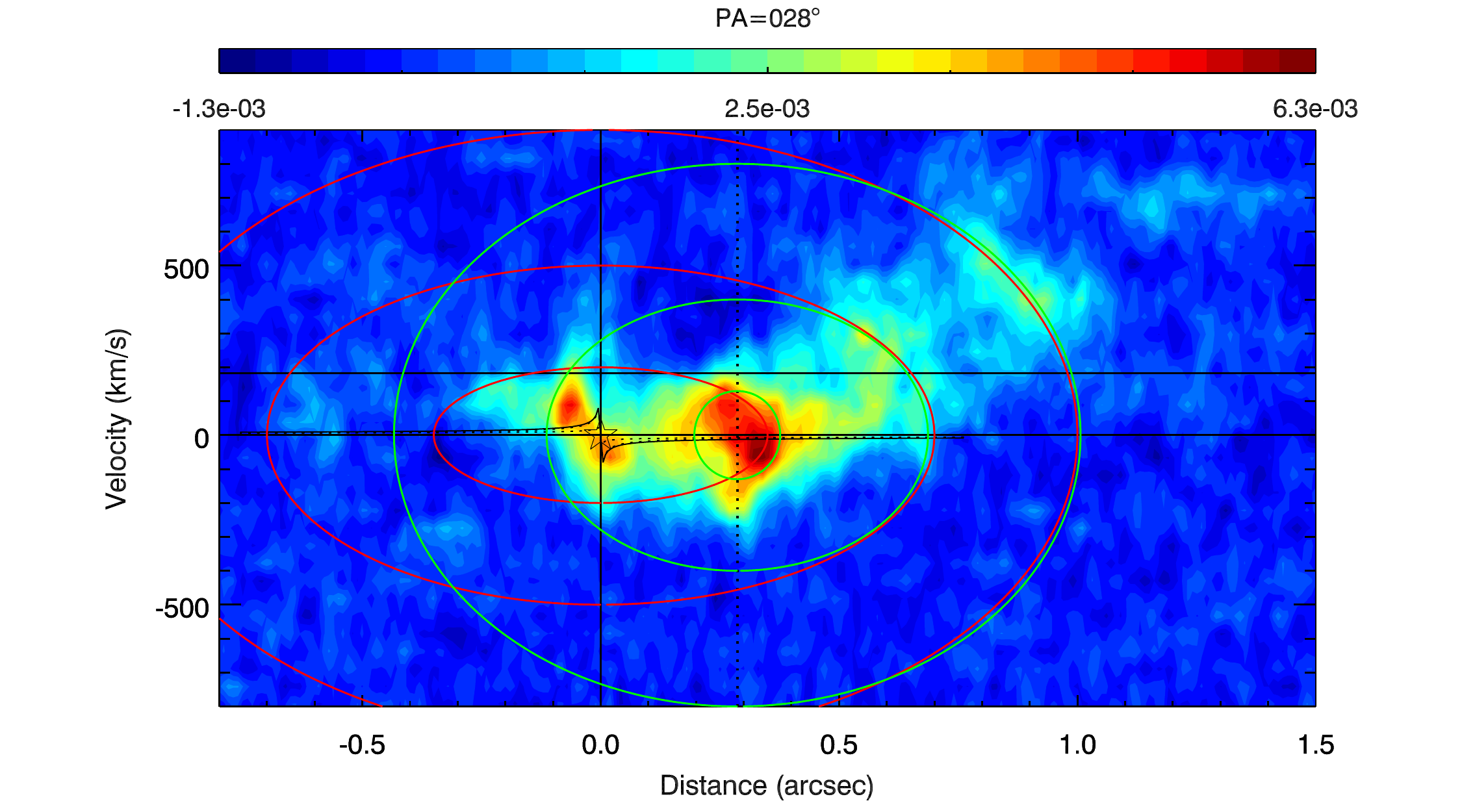}
\caption{Position-velocity diagram for the $^{12}$CO(2-1) emission along a PA of 28\arcdeg, centered on the southern nucleus in position and velocity, and passing through the radio supernova RS2 (at offset 0.3$''$).
To guide the eye, horizontal lines delineate the recessional velocities of the N and S nuclei and vertical lines delineate the positions of the S nucleus and RS2.  Also to guide the eye, we show for the South nucleus the predictions of the velocity fields for the
galaxy-only ({\it dashed curve}), black hole only ({\it dashed curve}) and total (galaxy plus black hole) enclosed mass ({\it solid curve}; see text). Ellipses delineate the expected observed velocities of radially expanding spherical shells around the 
South nucleus (red; velocities 200, 500, and 900 km/s at radii 0\farcs35, 0\farcs7, and 1$''$, respectively)  and RS2 (green; velocities 130, 400, and 800 km/s at radii 0\farcs09, 0\farcs4, and 0\farcs72, respectively).\label{alma_nuclear_pv028}}
\end{figure}

There is significant evidence for symmetric superwind-driven outflows from the South nucleus \citep{heckman90,tecza00,baan07}. Our high resolution molecular kinematics traces multiple outflow regions and shells. Figure~\ref{alma_nuclear_pv028} shows a p-v diagram which crosses both the
South nucleus and the location of the radio supernova RS2. As in Fig.~\ref{alma_nuclear_pv}, we clearly see high velocity dispersions around the location of RS2 (dotted vertical line). Further, even higher velocities are seen to the northeast of RS2 (larger positive offsets in the p-v diagram).
For illustration we have overplotted the expected observed velocities from expanding spherical shells of gas with expansion center at the Southern nucleus (red) and the RS2 position (green). These are illustrative `toy' models fitted by eye; the true picture is certainly more complex
as the molecular gas is likely swept up from the internuclear bridge, which is likely highly inclined to our line of sight (see \S1), and the shell morphology will depend strongly on the surrounding gas distribution with which it interacts.  Further, the significantly different velocities required to 
explain the different shells in our toy model either imply that the periodic outflows are weakening over time or that we have ignored projection factors and local systemic velocities (If the shells are comprised of gas swept up from the internuclear bridge then the inclination of this bridge to the line 
of sight and the fact that the bridge's systemic velocity gets more redshifted with larger offsets from the southern nucleus will both help to decrease the true difference in outflow velocities between shells. In any case, the dynamic time scales of these shells are significantly less than the estimated 
age of 10~Myr for the nuclear starburst \citep{tecza00}. As seen in Fig.~\ref{alma_nuclear_pv028}, the observed shells can be equally well explained as originating in either the Southern nucleus or (near) RS2. However, there are two arguments which support an origin of the outflows 
from a region close to RS2: (a) the innermost shell extends over more than 180\arcdeg\ in azimuth with a center very close to (but $\sim$0.02$''$ south of) RS2. If the outflow were driven by the South nucleus, we would expect the arc to be present only on the far side of RS2 (with respect 
to the Southern nucleus); (b) if the shells originated in the southern nucleus, one would expect to see high velocity arcs over a larger range of offsets  from the Southern nucleus. What we observe, however, is that these are found at offsets consistent with them being driven out of the gas rich 
region near RS2.  Given that RS2 likely traces a more extended starburst region, a center slightly offset from RS2 is not unexpected. 

The unusual kinematics in the sphere of influence of the South nucleus also deserves comment. Here, as in Fig.~\ref{alma_nuclear_pv}, the velocities within $\pm$0.1$''$ of the southern nucleus do not follow those expected from the rotating model. \citet{hagiwara11} detected a young 
radio supernova RS1 35~mas southwest of the south nucleus. We note that the velocity pattern seen near the South nucleus appears symmetric about a negative offset (consistent with a shift towards the position of RS1) and a positive recessional velocity (consistent with the kinematics of 
the internuclear bridge) with respect to the southern nucleus. A detailed analysis of the outflows driven by RS1, RS2, S, and N are deferred to a forthcoming work.

\begin{figure}[h!]
\plotone{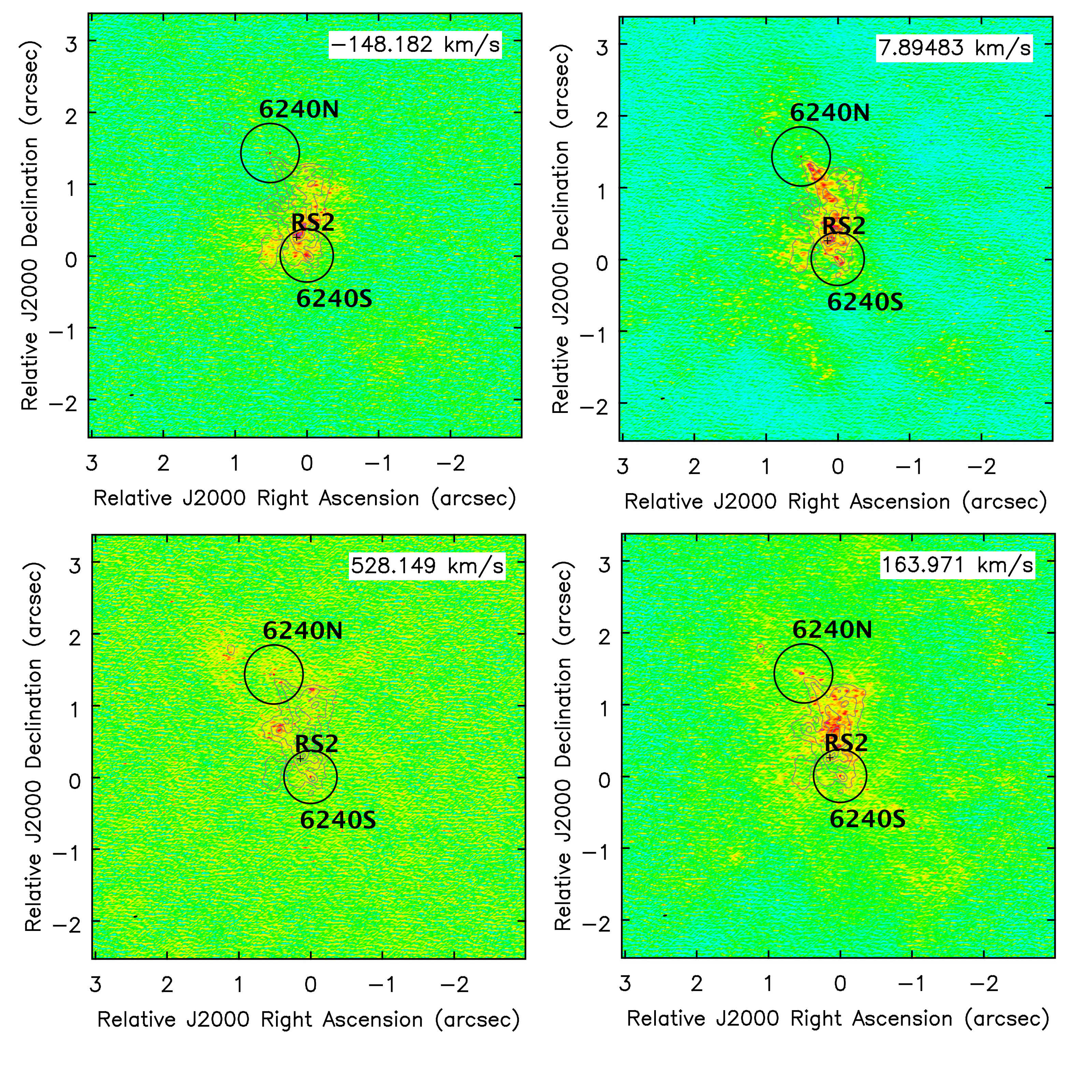}
\vspace{-0.5cm}
\caption{$^{12}$CO(2-1) emission as measured by ALMA in NGC 6240 in four channel maps: -148.2 km/s (blue-shifted stream; top left panel), 7.9 km/s (systemic; top right panel), 164.0 km/s (red-shifted stream; bottom right panel) 
and 528.1 km/s (outflowing material; bottom left panel). The gray contours in all panels show the integrated $^{12}$CO(2-1) emission, as shown in Figure~\ref{ngc6240_alma_hst}. Each contour shows an increase of 20\% in flux starting from
a base level of 0.2 Jy/beam km/s. The black circles have radii of $\sim$200 pc, centered on the position of the VLBI emission. \label{alma_co21_velcuts}}
\end{figure}

We now analyze the kinematic properties of the molecular gas at larger scales, by coarsely sampling the ALMA $^{12}$CO(2-1) spectral window in 26 km/s bins. Figure~\ref{alma_co21_velcuts} shows the detected $^{12}$CO(2-1) emission at four specific velocities relative to the systemic 
velocity of the system, defined to be that of the southern nucleus: -148.2 km/s (blue-shifted stream), 7.9 km/s (systemic), 164.0 km/s (red-shifted stream) and 528.1 km/s (outflowing material). Most of the blueshifted material is found near the southern part of the stream, while 
the redshifted emission is seen near the northern end. This suggests that the stream can be considered as a dynamic structure that is interacting with and evolving relative to the two nuclei.

\subsection{Velocity Dispersion}

\begin{figure}[h!]
\plotone{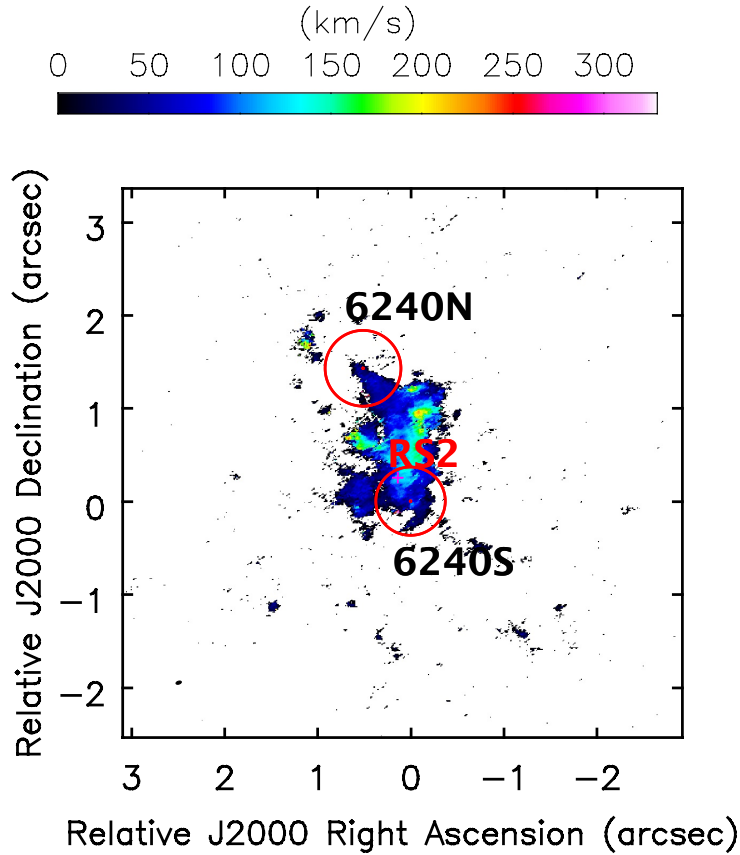}
\vspace{-7cm}
\caption{Velocity dispersion map for the molecular gas in NGC 6240, as traced by ALMA using the $^{12}$CO(2-1) emission line. The red circles have radii of $\sim$230 pc, roughly the sphere of influence for each SMBH, centered on the position of the 
VLBI emission (magenta dots), while the magenta cross marks the position of RS2. \label{alma_co21_veldisp}}
\end{figure}

The velocity dispersion, in particular relative to the kinematical structure of the system, provides important clues about the dynamical state of the molecular gas in the nuclear regions of this system. Previous observations, such as those presented 
by \citet{feruglio13}, found that at lower, $\sim$1$''$, spatial resolutions, velocity dispersions fell within the range $\sim$150-300 km/s but reached up to $\sim$500 km/s in the region in between the nuclei and around the northern center. Our ALMA observations,
shown in Figure~\ref{alma_co21_veldisp}, find consistent results, with typical velocity dispersion values in the region between the nuclei of $\sim$130-150 km/s, and higher values, reaching $\sim$250 km/s, mostly along the edges of the inter-nuclear CO emission. In 
the northern nucleus , using a radius of $\sim$200 pc, we find an average velocity dispersion of 184 km/s, while for the southern one, on a similar area, we find an average velocity dispersion of 155 km/s. These values are significantly lower than those reported 
by \citet{feruglio13}, and highlight how the spatial resolution of the two observations, $\sim$1$''$ versus $\sim$0.03$''$, plays a crucial role in the interpretation of the gas state near the SMBHs. Indeed, Submillimeter Array (SMA) observations of the $^{12}$CO(3-2) line carried out 
by \citet{u11} at a resolution of $\sim$0.3$''$, similarly finds a modest velocity dispersion for the molecular gas, fully consistent with what is reported here. Comparable values were also found from ALMA observations of the [C I](1-0) line emission, reported by \citet{cicone18}.

Albeit lower than previously reported, these still relatively high velocity dispersion values imply $v$/$\sigma$$\sim$1, and hence suggest that the kinematics of the inter-nuclear gas is dominated by turbulence, possibly
associated with the presence of significant shocks in the inter-nuclear molecular gas. In addition, and as presented by \citet{u11}, the velocity dispersion distribution is not connected to the overall velocity gradient between the 
nuclei. This latter fact would indicate that the bulk of the molecular gas, traced by the CO emission, is not undergoing organized circular movements as proposed previously \citep[e.g.,][]{engel10}, but instead is consistent with being dynamically supported by turbulence. This 
seems natural if this is indeed a transient structure created by the merger, which after a relatively short time (compared to the duration of the merger) may (at least partially) collapse around the new coalesced center of the system, as was previously discussed 
by \citet{engel10} and others, and predicted by numerical simulations \citep{bournaud11}.

\section{Discussion}\label{sec:discussion}

\subsection{Gas Mass from $^{12}$CO(2-1)}

While H$_2$ is the most abundant molecule in galaxies, it cannot be directly observed in emission due to its lack of dipolar rotational transitions at relatively low excitation temperatures \citep[][and references therein]{bolatto13}. In contrast,
the CO molecule can be easily excited even in cold environments and is relatively abundant, making it an important tracer of the molecular gas contents of a galaxy. In order to convert from observed $^{12}$CO(2-1) line flux to total molecular gas mass 
we follow the procedure described by \citet{solomon05}, as done previously by \citet{treister18}. Specifically, the line luminosity is computed from the observed line flux as

\[
L'_{\textnormal{CO}}=3.25\times 10^7 S_{\textnormal{CO}}\Delta v \nu_{obs}^{-2}D_L^2 (1+z)^3.
\]

From the line luminosity we estimate the molecular hydrogen mass as

\[
M(H_2)=\alpha_{\textnormal{CO}} L'_{\textnormal{CO}}
\]

\noindent where $\alpha_{CO}$ is the CO-to-H$_2$ conversion factor. Then, by incorporating the He contribution into this factor, we obtain the total molecular gas mass. The value of $\alpha_{CO}$ and its possible variations for different galaxies and
even regions inside a galaxy are topics still extensively debated in the literature \citep[e.g.,][]{bolatto13}. The value typically assumed for galaxies like the Milky Way is 4.6 $M_\odot$(K km s$^{-1}$pc$^2$)$^{-1}$ \citep{bolatto13}. However, in extreme environments such as those 
observed in (U)LIRGs,  lower $\alpha_{CO}$ values of $\sim$0.6-1 $M_\odot$ (K km s$^{-1}$pc$^2$)$^{-1}$ have been reported \citep[e.g.,][]{downes98,tacconi08,yamashita17, herrero-illana19}, likely associated with higher velocity dispersions \citep{papadopoulos12}. 
For a source as complex as NGC 6240 it is dangerous to assume a single value of $\alpha_{CO}$, as previously argued by \citet{engel10}. Previously, for this galaxy, \citet{tunnard15} reported a very broad range of $\alpha_{CO}$ values spanning up to three orders of magnitude,
from $\sim$0.1 to $\sim$100 $M_\odot$ (K km s$^{-1}$pc$^2$)$^{-1}$ (see their Figure 10, left panel). However, the reported range in the global value of $\alpha_{CO}$ is much narrower, finding $\alpha_{CO}$=1.5$^{7.1}_{1.1}$ $M_\odot$ (K km s$^{-1}$pc$^2$)$^{-1}$.
In parallel, using combined observations of HCN, CS and HCO$^+$ transitions, \citet{papadopoulos14} estimated a conversion factor of $\alpha_{CO}$=2-4 $M_\odot$(K km s$^{-1}$pc$^2$)$^{-1}$. More recently, based on ALMA observations at lower spatial resolutions than 
those presented here, \citet{cicone18} reported a mean value of  $\alpha_{CO}$=3.2$\pm$1.8 $M_\odot$(K km s$^{-1}$pc$^2$)$^{-1}$ for the systemic components and $\alpha_{CO}$=2.1$\pm$1.2 $M_\odot$(K km s$^{-1}$pc$^2$)$^{-1}$ for the outflowing material (presented in 
their Table 2). We adopt the \citet{cicone18} $\alpha_{CO}$ conversion factor for our calculations, together with their values of $r_{21}$=1.0$\pm$0.2 for the systemic component and 1.4$\pm$0.3 for the outflows, where $r_{21}$ is defined as the ratio between the $^{12}$CO(2-1) 
and $^{12}$CO(1-0) luminosities. When comparing to previous results, it is important to consider that fiducial values of $\alpha_{CO}$1~$M_\odot$(K km s$^{-1}$pc$^2$)$^{-1}$ were typically assumed \citep[e.g.,][]{engel10}.

We adopt three distinct regions: The spheres of influence of the northern and southern SMBHs, and a 1$''$ diameter aperture centered equidistantly along the line connecting the two nuclei (16$^h$52$^m$58.896, 02$^o$24$'$03.96$''$). The velocity-integrated $^{12}$CO(2-1) fluxes 
measurements are 34.3$\pm$3.7 Jy km/s, 142.1$\pm$3.1 Jy km/s and 389$\pm$11 Jy km/s, respectively. The central 1$''$ flux density is similar, albeit $\sim$25\% smaller, to the one reported by \citet{tacconi99}, which might indicate that our interferometric observations 
may have resolved out a small fraction of the inter-nuclear molecular mass; however, the lower resolution value of \citet{tacconi99} likely includes substantial flux from the nuclei, at least partially accounting for the 25\% difference.  The fluxes for the two nuclei are fully 
consistent with those measured previously by \citet{engel10}. As presented in section~\ref{sec:results}, considering a much larger aperture of 25$''$ in diameter, that should include most of the source $^{12}$CO(2-1) emission, we measure a 
flux of 1163$\pm$28 Jy km/s, consistent with the 1220 Jy km/s reported by \citet{tacconi99} and the $\sim$1000 Jy km/s inferred from the $^{12}$CO(1-0) 
observations by \citet{solomon97}. 

Using the expressions derived above, these line fluxes correspond to gas masses of 7.4$\times$10$^8$M$_\odot$ for the northern nucleus, 
3.3$\times$10$^9$M$_\odot$ for the southern nucleus and 8.6$\times$10$^9$M$_\odot$ for the central region. This is consistent with the values previously 
derived by \citet{engel10}, and \citet{feruglio13}, once the difference in the assumed values of $\alpha_{CO}$ are accounted for. Finally, from the computed 
total flux in the 25$''$ diameter aperture we derive a mass of 2.6$\times$10$^{10}$M$_\odot$. This is fully consistent with the value of 
2.1$\pm$0.5$\times$10$^{10}$M$_\odot$ reported by \citet{cicone18} in a central 12$''$$\times$6$''$ region. 

The total molecular gas mass can be also estimated from the warm molecular gas, as presented by \citet{dale05}, even considering that the latter
is not a direct tracer of the cold molecular gas. Specifically, for our comparison we can use the H$_2$1-0 S(1) line luminosity at rest-frame 2.12~$\mu$m. We use the empirically-calibrated relation L$_{\textnormal{(1-0)S(1)}}/M_\textnormal{gas}$ = 2.5$\times$10$^{-3}$ \edit1{as measured specifically for NGC 6240 by} \citet{muller-sanchez06}.  This luminosity can be estimated from the VLT/SINFONI observations 
of NGC 6240 presented by \citet{engel10}. Using a nuclear 2$''$ diameter aperture we estimate a L$_{\textnormal{(1-0)S(1)}}$ luminosity of 2.8$\times$10$^7$L$_\odot$, or equivalently 1.1$\times$10$^{10}$M$_\odot$, and hence fully consistent with the values derived here based on the 
$^{12}$CO(2-1) observations. The comparison of the results obtained using the empirically-calibrated
L$_{\textnormal{(1-0)S(1)}}/M_\textnormal{gas}$ ratio with the new ALMA observations presented in this paper indicates that we can use the H$_2$1-0 S(1)
line to make at least an approximate estimate of the total (cold) molecular gas mass.

\subsection{Gas Mass from sub-mm continuum}

In the past,  most studies of the ISM in the molecular gas phase have relied on the rotational transitions of the CO molecule to trace H$_2$ gas. However, as shown by \citet{scoville14,scoville16}, measurements of the dust continuum emission at the long 
wavelength Rayleigh-Jeans (RJ) end can also provide accurate estimates of the inter-stellar medium (ISM) mass in a small fraction of the required observing time with ALMA. At these long wavelengths, the dust emission is thought to be optically thin and hence
the observed continuum flux density is directly proportional to the mass, once the dust opacity coefficient and the dust temperature are established. It is important to point out that this estimation depends on the assumption that the continuum emission at
these wavelengths is dominated by thermal dust radiation, which might not be the case in the central region of NGC 6240. Evidence in this direction was presented by \citet{tacconi99} and later by \citet{nakanishi05}, based on the observed spectral slope, which is consistent
with that expected for synchrotron radiation. In addition, \citet{meijerink13} argued that the CO to continuum luminosity ratio in this galaxy is significantly larger than the value observed in other similar systems, which might be an indication of significant shock heating. However,
as it was presented by \citet{medling19}, the estimated synchrotron emission in the northern and southern nuclei are 0.16 and 0.08 mJy, respectively, and hence negligible compared to the observed continuum fluxes, reported in Section 3.2. Therefore, we can assume that
the continuum emission is dominated by thermal emission from dust and hence can compare this estimate to the molecular gas mass derived in the previous section. Following the procedure presented by \citet{scoville15}, 
the ISM mass is given by their equation 3:

\[
M_{\textnormal{ISM}}=\frac{0.868\times S_\nu (mJy)d^2_{Gpc}}{(1+z)^{4.8}T_{25}\nu^{3.8}_{350}\Gamma_{RJ}}10^{10}M_\odot,
\]

\noindent where $\Gamma_{RJ}$ is a correction factor that accounts for the departure from the RJ regime and is given by

\[
\Gamma_{RJ}=0.672\times \frac{\nu_{350}(1+z)}{T_{25}}\times\frac{1}{e^{0.672\times\nu_{350}(1+z)/T_{25}}-1}.
\]

For our observations of NGC 6240, $\nu_{obs}$ corresponds to 235 GHz, or 0.67 in $\nu_{350}$ units, $z$=0.02448 and $d_{Gpc}$=0.1. Furthermore, we assume $T_{25}$=1. From the ALMA compact and intermediate configurations band 6 continuum maps of the southern and northern nuclei, as defined 
by the radius of the SMBH sphere of influence, we measure 235 GHz continuum fluxes of 5.19$\pm$0.06 mJy and 1.73$\pm$0.03 mJy, respectively, 
as presented in section~\ref{sec:continuum}. These correspond to ISM masses of 2.1$\times$10$^9M_\odot$ and 6.9$\times$$10^8M_\odot$
respectively. These masses are roughly consistent with those derived from the CO measurements in the previous section. Differences of that order are 
expected given the typical uncertainties in the assumed values of $\alpha_{CO}$ and $T_{25}$, that cannot be determined directly from our ALMA data at these high resolutions. 

To determine  if the difference in molecular mass estimations could be caused by a surface brightness sensitivity issue, we use the \citet{scoville16} relation to calculate the expected $850~\mu m$ luminosity from the 
CO-derived molecular mass within the central 2\arcsec~aperture. We then convert this luminosity to an equivalent 235 GHz continuum emission (again 
assuming $\beta=1.8$) and estimate the flux density per beam in our long-baseline observations. We find that the predicted dust continuum emission per beam associated with 
the CO-detected molecular concentration in the inter-nuclear 1$''$ region is an order of magnitude fainter than the sensitivity of our observations. This indicates that dust-based 
ISM mass estimates can be biased low in situations where galaxies are highly resolved.

\subsection{Outflowing Material}
\label{sec:outflow}

\begin{figure}
\begin{center}
\plotone{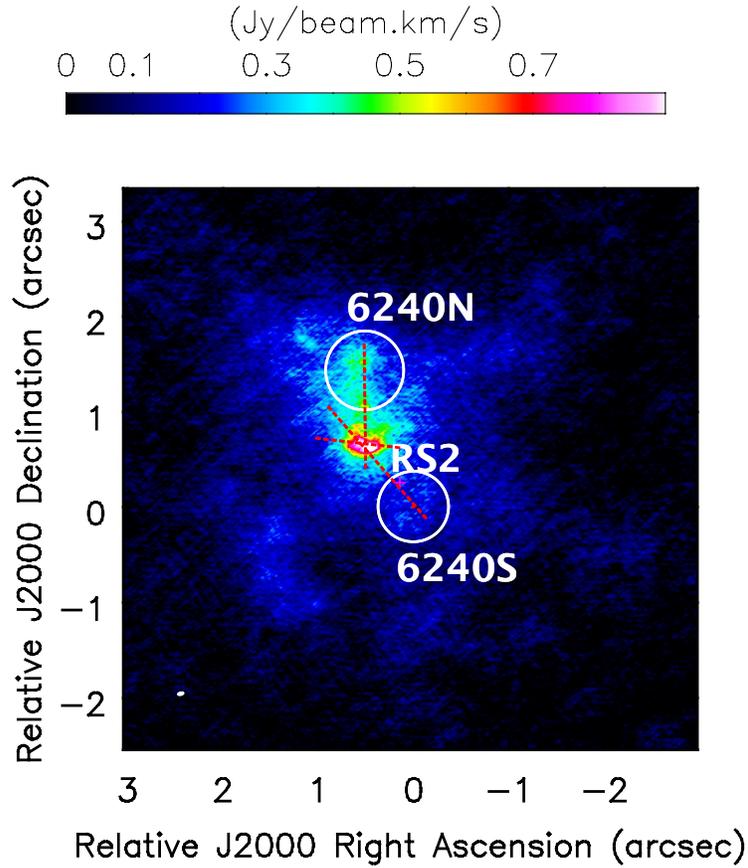}
\end{center}
\vspace{-7cm}
\caption{ALMA moment 0 map of the $^{12}$CO(2-1) emission at relative velocities greater than 500 km/s. Symbols are the same as in Figure~\ref{alma_velmap}. A clear high velocity emission concentration is visible directly to the south of the northern 
nucleus. This emission appears to be connected to the northern nucleus by a fainter emission bridge. Three position-velocity diagrams, presented in Figure~\ref{alma_outflow_PV}, were extracted to characterize the high velocity structure, as shown by the dotted 
red lines.\label{alma_outflow}}
\end{figure}

The velocity map presented in Figure~\ref{alma_velmap} shows a high velocity, $>$500 km/s, component separated by $\sim$400 pc in projection to the south of the northern nucleus. A map showing the structure of this high velocity material is presented in Figure~\ref{alma_outflow}. There 
seems to be a faint bridge of gas extending back to the northern nucleus. This structure was also presented and studied by \citet{feruglio13}, and later by \citet{saito18} from lower resolution $^{12}$CO(2-1) observations. More recently, \citet{cicone18} reported the detection of this
nuclear outflow in the CO(1-0), CO(2-1) and [C I](1-0) transitions, and \citet{mueller-sanchez18} in $H_2$. Given its high velocity, we speculate that this is an outflow of molecular gas expelled from the nuclear region of the merger system. The total mass of this outflowing material is 
estimated at 9$\times$10$^8$ $M_\odot$ considering  the $\alpha_{CO}$ factor computed specifically for the outflowing material by \citet{cicone18}. This is very significant, as it represents $\sim$3.5\% of the total molecular gas in the system and 10.5\% of the molecular mass in the central 
region. If we assume that it is linked to the northern nucleus and has been moving at constant speed, this material would have been expelled at least 0.78 million years ago. 

We estimate the mass outflow rate associated with this high velocity structure using equation 3 in \citet{maiolino12}:
\[
\dot{M}_{\textnormal{out}}=3v\frac{M_{\textnormal{out}}}{R_{\textnormal{out}}},
\]
where $v$ is a representative velocity of the wind, $M_{\textnormal{out}}$ its mass and $R_{\textnormal{out}}$ its size. This equation assumes a relatively
simple geometry for the outflow, given by a spherical, or equivalently conical or multi-conical, volume uniformly filled by outflowing clouds. We caution that the geometry of this outflow is likely more complex and hence the values derived using this expression have
uncertainties of factors of $\sim$3 \citep{lutz19}. Assuming values of $v$=500 km/s, $M_{\textnormal{out}}$=9$\times$10$^8$ $M_\odot$ and $R_{\textnormal{out}}$ between 
290pc and 430pc depending on whether the wind arises from either the northern nucleus for the latter or the extension of the high velocity structure for the former. We then obtain values of 3200 M$_\odot$/yr to 4700 M$_\odot$/y, roughly consistent with the value of 
$2500\pm1200$ M$_{\odot}$/yr reported by \citet{cicone18}.

We can then constrain the ultimate fate of the wind by comparing the wind velocity with the system's escape velocity. NGC~6240 has a stellar mass of $3.9\times10^{11}$ M$_{\odot}$ \citep{howell10}; with the abundance matching of \citet{moster10} the predicted halo mass 
is $\sim12\times10^{14}$ M$_{\odot}$. The escape velocity for a halo of this mass is $\sim1000$ km~s$^{-1}$, computed at $R_{200}$ ($\sim1$ Mpc). Assuming there is no further injection of energy, most of the mass will remain bound to the halo.
If the material in the wind is efficiently heated and joins the X-ray halo \citep{nardini13}, future star formation may be delayed or prevented as a result of the wind's removal of material from the center of NGC~6240. The presence of this high-speed massive clump implies a 
very efficient process capable of accelerating high-velocity massive amounts of material and consequently, due to the gas reservoir removal, significantly diminishing or even shutting down nuclear star formation episodes triggered 
by the major galaxy merger. Taking the total molecular gas mass of $2.6\times10^{10}$ M$_{\odot}$ previously derived and the outflow rate of 3200~M$_{\odot}$ yr$^{-1}$ the implied gas removal time by the wind alone is 8.1 Myr.

\begin{figure}
\begin{center}
\includegraphics[scale=0.45]{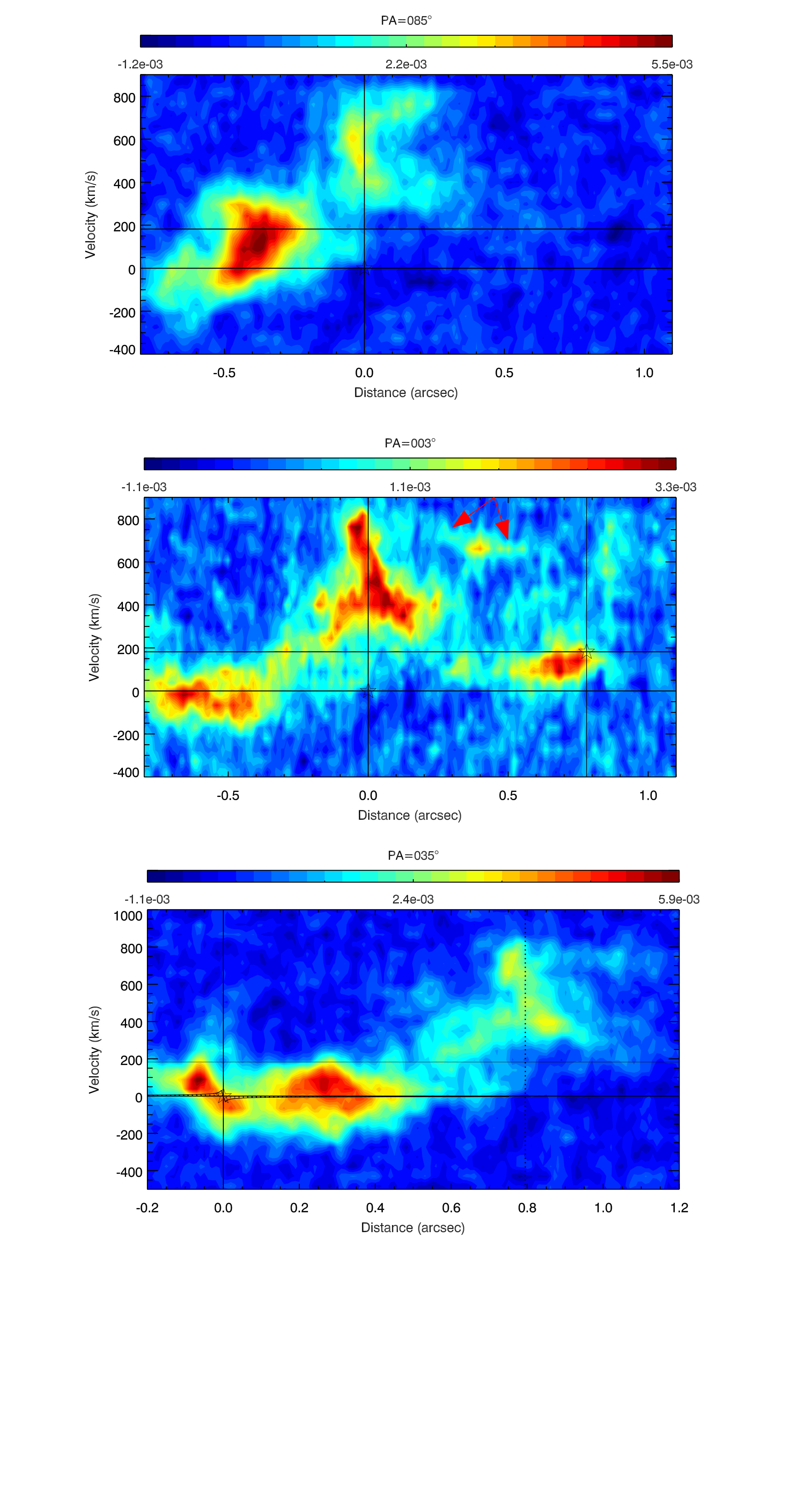}
\end{center}
\vspace{-4.5cm}
\caption{Position-velocity diagrams, specifically tracing the high velocity emission shown in Figure~\ref{alma_outflow}. The {\it top panel} shows a diagram along the brightest axis of the emission, at a PA of 85\arcdeg, roughly aligned in the east-west direction, while the 
{\it center panel} shows a cut along the fainter material connecting the northern nucleus to the center of the high velocity emission, roughly in the north-south direction, at a PA of 3.3\arcdeg. The {\it bottom panel} presents the p-v diagram from the high velocity 
emission to the southern nucleus, at a PA of 35\arcdeg. In the top and center panels the vertical line at an offset of 0$''$ marks the center of the high velocity emission, while in the bottom panel the southern nucleus is at an offset of 0$''$ and the high velocity emission at $\sim$0.8$''$. No 
well defined kinematic structure can be seen in the region of the high velocity emission. The faint emission connecting the high velocity structure to the northern nucleus seen in Fig.~\ref{alma_outflow} is visible on the central panel, at a velocity of $\sim$600 km/s and offsets 
between 0.2$''$ to 0.6$''$, indicated by the red arrows. \label{alma_outflow_PV}}
\end{figure}

In order to explore the kinematic properties of the aforementioned high velocity material, in Figure~\ref{alma_outflow_PV} we present three p-v diagrams, one along the bulk of the high velocity emission, roughly in the east-west direction, another tracing the material bridging this
cloud to the northern nucleus, roughly oriented north-south, and another connecting the high velocity emission to the southern nucleus, as shown in Figure~\ref{alma_outflow}. No evidence for rotation or obvious kinematic structures can be seen in the high velocity emission.
This emission extends from $\sim$300 km/s to $\sim$800 km/s, spanning $\sim$0.3$''$ in size, which correspond to $\sim$150 parsec in projection at the distance of NGC 6240. The north-south position-velocity diagram (middle panel in Figure~\ref{alma_outflow_PV}) shows a rather steep
gradient, which might be indicative for example of an expanding shell. A faint connecting bridge can be observed in Figure~\ref{alma_outflow} between the northern nucleus and the high velocity structure. This emission can also be seen in the central panel of Figure~\ref{alma_outflow_PV} at a
relatively constant velocity of $\sim$600 km/s and spanning spatial offsets between 0.2$''$ and 0.6$''$, almost reaching the northern nucleus. No similar connection can be seen relative to the southern nucleus (bottom panel in Fig.~\ref{alma_outflow_PV}). As reported by \citet{mueller-sanchez18}, the 
high velocity outflow is clearly detected in the $H_2$ VLT/SINFONI map and in spatially-coincident [OIII] emission, as observed in {\it HST} Wide Field Camera 3 (WFC3) maps of the region. Considering as well the lack of H$\alpha$ emission at this location, also reported by \citet{mueller-sanchez18},
the physical origin of this structure remains uncertain. Given its complexity, a full kinematic analysis of this high velocity structure and models for its origin are beyond the scope of this paper.

\subsection{Molecular Gas as Fuel for Star Formation}

As shown by \citet{medling15}, both SMBHs are currently more massive than expected based on the black hole mass-galaxy properties correlations, requiring an increase in the total stellar mass by 1.7$\times$10$^{11}$$M_\odot$ to reach the scaling relation. Assuming that 
all the molecular gas currently in the sphere of influence of each SMBH will be accreted by them, the SMBHs would only grow by $\sim$45\%, and hence not change too much in overall mass in the short term. The total mass reservoir is however significantly larger, considering 
the $\sim$9$\times$10$^9$M$_\odot$ of molecular gas in the central 1$''$ (500pc) diameter inferred from the ALMA  $^{12}$CO(2-1) observations. Its precise fate, if to be accreted or to form stars or to be expelled from the system through outflows, is highly uncertain at this point. However, 
recent observations of a similar system, Mrk 463 \citep{treister18}, show that only a very small fraction, $<$0.01\%, of the available mass is actually being accreted by the SMBHs. Therefore, we could expect that, after subtracting the outflowing material, most of the remaining molecular gas will 
be used as fuel for star formation.  Assuming typical efficiencies for star formation for giant molecular clouds of $\sim$1-10\% \citep{murray11,ochsendorf17}, this implies an expected increase of the stellar mass in the nuclear region of (0.3-3)$\times$10$^8$M$_\odot$, which is certainly 
not enough to make the resulting system consistent with the observed black hole-stellar mass correlation. Even if all the molecular gas in the system is considered, these conclusions are not altered significantly, even if much higher star formation efficiencies are invoked. 

\section{Conclusions}\label{sec:conclusions}

The unprecedented high resolution ALMA observations of the molecular gas in the merging system NGC 6240 allow us to pin down how SMBH growth and star formation are proceeding in this complex
and chaotic environment. While previous observations of these and other nearby major galaxy mergers reached spatial resolutions of 100s-1000s pc, the recent availability of ALMA configurations reaching $>$10 km baselines allowed to sharpen this
view by an order of magnitude. This allows to resolve scales comparable to the sizes of average giant molecular clouds \citep{murray11}, and hence to measure size, mass and kinematics for many active sites of nuclear star formation. Therefore, and as we have shown here, we can now
study the properties of the gas inside the sphere of influence of each SMBH, which is readily available to feed them, quantify the general kinematic structure of the gas, and identify high velocity outflows. 

Specifically for NGC 6240, the high resolution band 6 ALMA observations of the $^{12}$CO(2-1) emission have confirmed that the bulk of the nuclear molecular gas, $\sim$9$\times$10$^9$$M_\odot$, is located in an $\sim$1$''$ (470 pc) region between the two nuclei. Significant 
amounts of molecular gas, 7.4$\times$10$^8$M$_\odot$ and 3.3$\times$10$^9$M$_\odot$ in the northern and southern nuclei, respectively, are found inside the sphere of influence of each SMBH. Contrary to previous, lower spatial resolution
observations, we do not find evidence for a kinematic rotating disk between the two nuclei. Instead, we find two spatially-localized velocity gradients reaching $\sim$200 and $\sim$400 km/s, both in the surroundings of the southern nucleus. We further find evidence for a very significant, $\sim$11\% of
the nuclear molecular mass, high velocity, $>$500 km/s, outflow. Its origin is not clear, but could be connected and physically linked to the northern nucleus. Studying the velocity dispersion, we find that the general dynamics of the molecular gas are consistent with
being dominated by turbulent or disordered motions and high velocity winds, most likely related to the ongoing major galaxy merger and nuclear activity. While the final fate of this molecular mass cannot be determined, it is expected that a fraction of it will eventually feed 
both SMBHs, which are currently above the well-established black hole mass-stellar mass correlation. This offset will not disappear even if a significant fraction of the molecular mass in the nuclear region is converted into stars at typical efficiencies. Major galaxy mergers are relatively rare, particularly 
in the local universe, with only a few tens close enough and bright enough to carry out such sensitive high resolution observations. It is particularly important to be able to carry out this high resolution analysis for a sample of sources spanning different merger stages, in order to be able to establish 
the evolution of the molecular gas across the merging sequence, a process that should take on average 1-2 Gyrs.

\acknowledgments 

We thank the anonymous referee for a very constructive review of this article. We acknowledge support from FONDECYT Regular 1160999 (ET), 1190818 (ET, FEB), CONICYT PIA ACT172033 (ET, FEB), Basal-CATA PFB-06/2007 (ET, FEB) and AFB170002 (ET, FEB, GV) grants,
and Chile's Ministry of Economy, Development, and Tourism's Millenium Science Initiative through grant IC120009, awarded to The Millenium Institute of Astrophysics, MAS (FEB). GCP acknowledges support from the University of Florida. CC acknowledges funding from the European 
Union Horizon 2020 research and innovation programme under the Marie Sk􏰁lodowska Curie grant agreement No.664931. This work was performed in part at the Aspen Center for Physics, which is supported by National Science Foundation grant PHY-1607611. This paper makes use of the 
following ALMA data: ADS/JAO.ALMA\#2015.1.00370.S and \#2015.1.00003.S. ALMA is a partnership of ESO (representing its member states), NSF (USA) and NINS (Japan), together with NRC (Canada), MOST and ASIAA (Taiwan), and KASI (Republic of Korea), in cooperation with the 
Republic of Chile. The Joint ALMA Observatory is operated by ESO, AUI/NRAO and NAOJ. The National Radio Astronomy Observatory is a facility of the National Science Foundation operated under cooperative agreement by Associated Universities, Inc.

\vspace{5mm}
\facilities{ALMA}

\appendix

\section{ALMA Array Setup and Data Analysis}

\subsection{Compact and Intermediate configurations}
\label{compact_conf}
The initial calibration of the short-baseline ($\lesssim1\,$km) had some slight changes in EBs Xb4da9a/X69a and Xb5fdce/X79d. In the former, the spectral slope of J1550+0527, derived based on the Titan amplitude solutions, was considered in order to obtain an improved bandpass solution. In the 
latter, J1751+0939 was used as flux calibrator instead of Pallas, which showed problematic amplitude visibilities inducing a general overestimate of the sources' fluxes. Also, despite the short times on source (1--4\,min), a single self-calibration step considering phase solutions alone was adopted 
(no major improvement was found by considering the amplitude solutions). This was done by setting the solution interval parameter to 'inf', and not adopting
any scan average. Since each scan was $\sim$50-60 seconds long, in practice this was the time range used.

\begin{figure}[h!]
\plotone{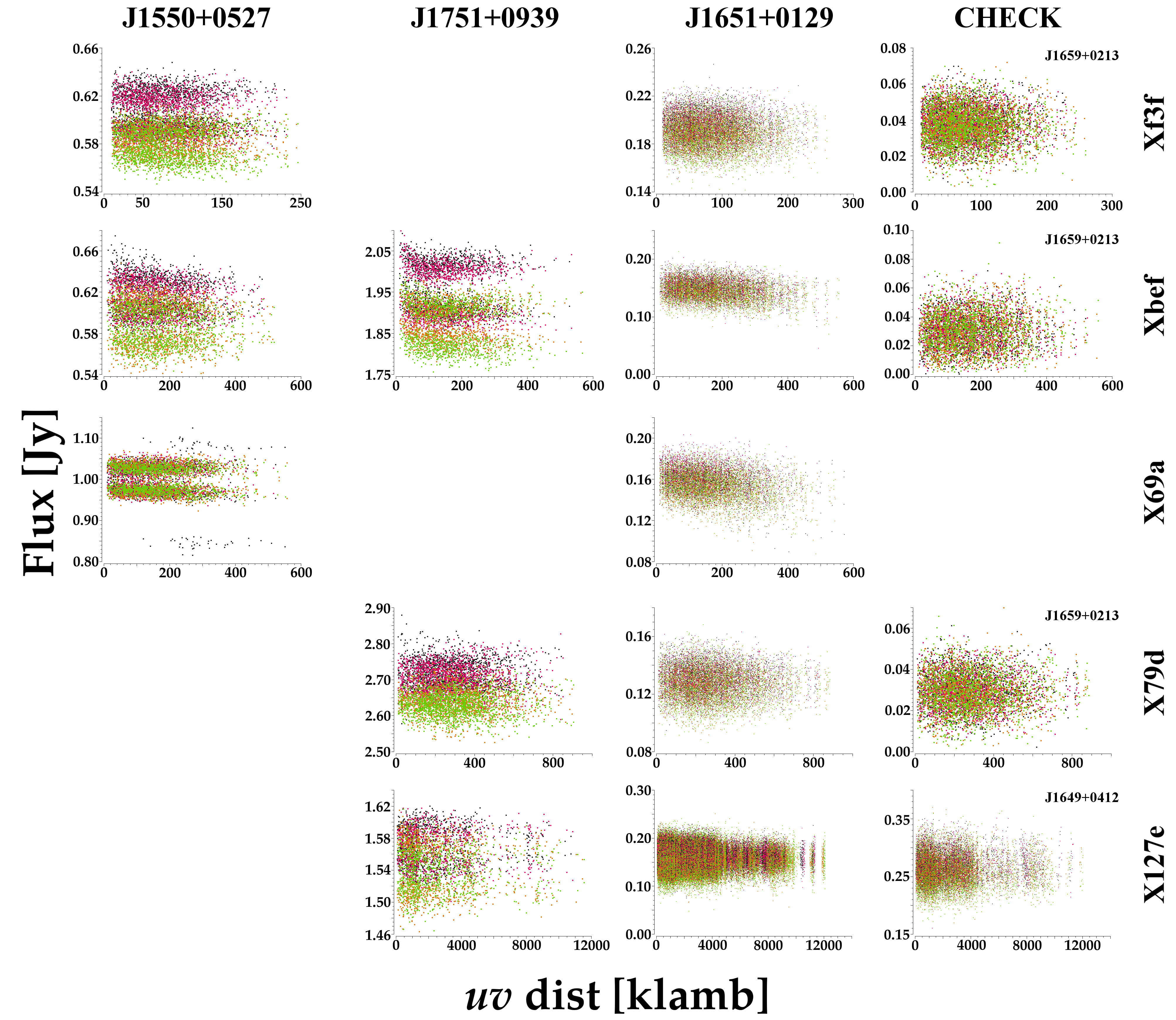}
		\caption{The amplitude versus \textit{uv}-distance visibility distributions obtained after the WVR solution scaling for the point-like sources used as calibrators in the observations considered in this work. \label{reduc_ampcals}}
\end{figure}

Figure~\ref{reduc_ampcals} shows the amplitude versus \textit{uv}-distance visibility distribution for the point-like calibrators used for flux, bandpass, or phase calibration, as well as the check sources. One can see that J1751$+$0939 shows modest large scale structure being picked up by the 
shortest baselines ($\lesssim50\,$k$\lambda$), but this difference is about 1\% above the core emission and thus is not expected to affect the calibration. Both J1550$+$0527 and J1751$+$0939 show visible variability either in flux, slope, or polarization. Comparing to the 
{\sc AMAPOLA}\footnote{\url{http://www.alma.cl/~skameno/AMAPOLA/}} data-base, which makes use of ALMA's Grid Survey observations, one can see that, J1550$+$0527 shows flux variability at the 10 to 18\% level from January to June 2016, and an increase of polarized flux in the same 
period. As for J1751$+$0939, there is an even larger variation in the flux and polarization fractions during the period of observations, with a peak in the first half of July 2016 (Xb5fdce/X79d). The fact that J1751$+$0939 was used as a flux calibrator in the execution Xb5fdce/X79d, which shows 
the highest flux density, does not affect the calibration. In fact, both J1651$+$0129 and J1659$+$0213, whose amplitude and bandpass calibrations depend on the previous sources, do not show the same flux, and polarization fraction variations nor any correlation. This indicates that the variation 
in J1751+0939 on that day was due to intrinsic source variability.

\begin{figure}[h!]
\plotone{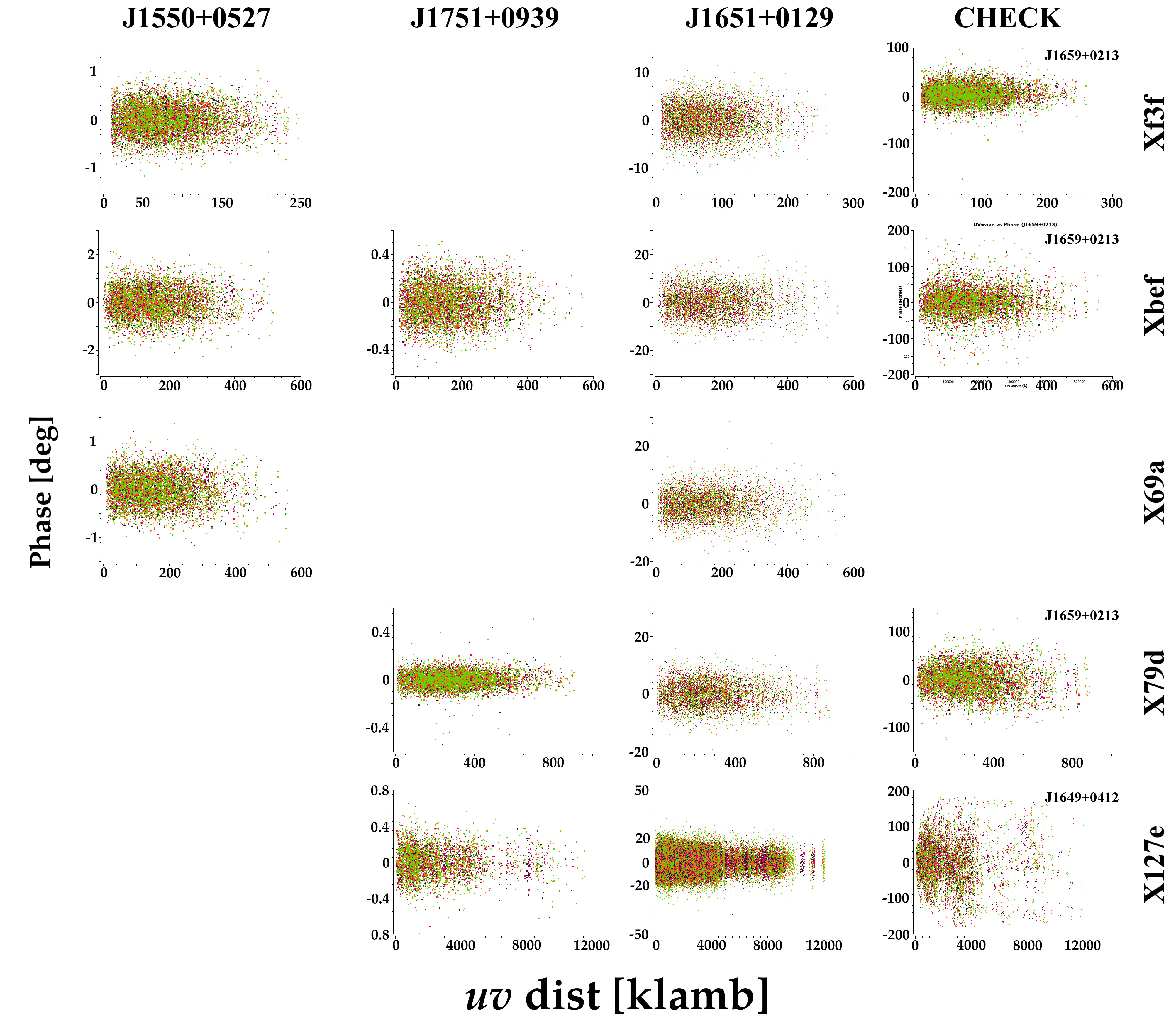}
	\caption{The phase versus \textit{uv}-distance visibility distributions obtained after the WVR solution scaling for the point-like sources used as calibrators 
	in the observations considered in this work. \label{reduc_phcals}}
\end{figure}

Likewise, we also display in Figure~\ref{reduc_phcals} the phase versus \textit{uv}-distance for the same sources. All the sources in these configurations show scatter around zero degrees, with no obvious trend with increasing baseline length, demonstrating that these sources are point-like 
and have been observed at the phase-centre. J1550+0527 and J1751+0939 show a phase scatter always below 1\,deg, while for the phase-calibrator J1651+0129 always appears below 10\,deg, implying that the signal decorrelation is smaller than 
1.5\% ($\epsilon\equiv e^{-\sigma_\phi^2/2}$, where $\sigma_\phi$ is the phase-rms in radians).

\subsubsection{High resolution}

Two visits were executed with the longest-baseline configuration. Both used J1651+0129 (1.02\,deg away) as phase-calibrator. The 2015 (Xac5575/X8a5f) visit was executed under poor phase stability ($\sigma_\phi\sim47\,$deg), resulting in  
strong signal decorrelation, while the second one in 2017 (Xc5148b/X127e) had better phase-stability ($\sigma_\phi\sim13\,$deg), even though still marginal (phase-rms at longer baselines of 40-70\,deg). This could be concluded from the dynamic range (DR) obtained in the 
continuum image of our scientific target, NGC 6240, resulting from the two executions. Although both observations have comparable noise levels ($\sim$29\,$\mu$Jy/beam), we find peak fluxes of the southern nucleus of 673\,$\mu$Jy (DR=24) and 2984\,$\mu$Jy (DR=102) in 
Xac5575/X8a5f and Xc5148b/X127e, respectively. Assuming this nucleus presented no significant flux variability within the time separating the two observations, this signal decorrelation is quite critical.

For this reason, we considered the use of the algorithm developed by \citet{maud17} which, based on the bandpass scan, finds the best WVR solution scaling to account for the dry-component in the atmosphere. We note that one should expect small changes since these 
observations were conducted under PWV$\sim$1, the upper limit where this scaling is expected to induce significant changes. The assessment of its application was done in two ways: determining the dynamic range of the NGC\,6240 continuum image (dominated by the 
two point-like nuclei) and the improvement of the phase-rms in each antenna in the phase-calibration and check-sources scans. When applying the scaling to these data-sets, we retrieve noise levels of 24 and 28\,$\mu$Jy/beam for Xac5575/X8a5f and Xc5148b/X127e, respectively. The 
peaks are now found to be 805\,$\mu$Jy/beam (DR=33; 5.5$\sigma$ increase) and 3031\,$\mu$Jy/beam (DR=108; 1.7$\sigma$ increase), respectively. The DR increase in the latter is small, but we decided to keep this scaling since the phase-rms improved for most antennas (41 out of 43 in the 
phase-calibrator scans, and 31 out of 43 in the check-source scans). Given the significant difference between the two data-sets, the final images consider the Xc5148b/X127e alone, which is less affected by signal decorrelation.

Further assessing the data-quality in Xc5148b/X127e, we focus on the check-source (J1649+0412). This is expected to be point-like, but after applying the phase-solutions based on the phase-calibration (J1651+0129, 2.74\,deg away), it appears to be slightly extended. The 
integrated-to-peak flux ratio is $\sim$1.5 (190 versus 130\,mJy) and the phase-rms is $\sim$40\,deg ($\epsilon\sim0.78$, Figure~\ref{reduc_phcals}). This may imply that we have bad phase solutions, in agreement with the large phase spread in Figure~\ref{reduc_phcals}. To address this, we 
self-calibrated the check-source data once (a single instance) assuming a point-like source model at the phase-centre. The solution interval was set per integration, and no solution was rejected at the 3$\sigma$ level. Table~\ref{tab:chksrc} summarizes the results assuming different pre-calibration 
strategies. As one can see, depending on whether the peak- or integrated-flux ratios are considered, there is a signal decorrelation of 25 to 50\% (the smaller value being in agreement with the observed phase scatter of $\sim$40\,deg). After self-calibration, we measure for the check source
a flux of $\sim$260\,mJy, compared to initial peak and integrated-fluxes of 130 or 190\,mJy, respectively. This relatively weak phase stability, combined with the sparse \textit{uv} coverage for long baselines affect particularly continuum measurements at scales of $\sim$1$''$, as presented
in \S\ref{sec:continuum}, but the $^{12}$CO(2-1) analysis at a much lower (negligible) level, since this was done using a much narrower frequency range, based on a feathering process on a per-channel basis.

\begin{sidewaystable}
	\centering
	\begin{tabular}{cccccccc} \hline
		WVRsc & Flag Ant's & peak & $1-\epsilon$ & integ & $1-\epsilon$ & integ/peak & off \\\relax
		[Y/N] &  & [mJy] & [\%] & [mJy] & [\%] &  & [mas] \\
		\hline
		Y & DA61 & 132$\pm$4 &  & 190$\pm$10 &  & 1.4$\pm$0.1 & 14\\
		&  & 261.9$\pm$0.2 & 50$\pm$1 & 261.3$\pm$0.3 & 28$\pm$2 & 0.998$\pm$0.001 & 0\\
		\hline
		N & DA61 & 132$\pm$5 &  & 190$\pm$10 &  & 1.5$\pm$0.1 & 13\\
		&  & 261.9$\pm$0.2 & 50$\pm$1 & 261.3$\pm$0.3 & 27$\pm$2 & 0.998$\pm$0.001 & 0\\
		\hline
		Y & high-phase & 167$\pm$4 &  & 196$\pm$9 &  & 1.18$\pm$0.06 & 15\\
		&  & 262.1$\pm$0.2 & 36.3$\pm$0.6& 262.2$\pm$0.4 & 25$\pm$1 & 1.000$\pm$0.002 & 0\\
		\hline
	\end{tabular}
	\caption{The self-calibration (selfcal) results on the check-source, for different pre-calibration strategies. The first and second sets of rows consider WVR scaling being applied (Y) or not (N), respectively. The bottom set of rows shows the results when antennas with high 
	phase-rms ($\sigma_\phi>1\,$radian or 57\,deg) are flagged in addition to DA61 (see text for more details). In each set of two rows, the first and second refer to the pre- and post-selfcal observed values, respectively. The decorrelation factors (in \%) between the selfcal'ed and 
	not selfcal'ed are shown in the second row in each set. The last column shows the offset between the source and the phase centers in milli-arcseconds.}
	\label{tab:chksrc}
\end{sidewaystable}

\subsubsection{Self-calibrating Xc5148b/X127e}

The previous section shows that a significant level of signal decorrelation is affecting the data. This is inducing an overestimate of the total source flux at large scales, since much of the line flux is extended (i.e., integrated to peak fluxes ratio $>1$). For this reason, we 
have adopted a step-wise self-calibration approach in baseline length. We first used the compact- and intermediate-configuration datasets to image the source. These covered the baseline range of $<800\,$k$\lambda$ (Baseline lengths can be converted to km multiplying
by a factor of 0.00133), which partly overlaps with those in Xc5148b/X127e. We thus used the first model image (from non-extended configurations) to self-calibrate only those antennas in Xc5148b/X127e which have at least three baselines in the overlapping range. Note that these 
antennas have longer baselines between themselves, which means that after self-calibration, one can retrieve a higher resolution image making use of those baselines longer than the previous cut (in our case, $<800$ versus $<2400\,$k$\lambda$). This new image allows the 
calibration of other antennas at longer baselines. Such step-wise (``unfolding'') approach was considered using the following baseline cuts: 800, 2400, 4000, 8000\,k$\lambda$. Each self-calibration step only accepted per-antenna solutions when at least 3 baselines were available 
within the baseline-length cut and significance was greater than 1.5$\sigma$ (probability greater than 86.6\%). The solutions were found per scan and averaged over both polarizations ({\rm gaintype}~=~T), since no significant difference between XX and YY phase-solutions was 
found. Each step always achieved less than 10\% rejected solutions, with the last phase-calibration step at $<8000\,$k$\lambda$ achieving 9.1\% rejection of the data initially available for self-calibration (i.e., after the usual phase referencing calibration). The final calibration adopted 
WVR scaling, with DA61 flagged, and the ``unfolding'' self-calibration strategy applied.

\subsubsection{Imaging Parameters}

Once all the data reduction and calibrations steps described above were performed, we carried out the imaging process in order to create the final data cube. This was done using a natural weighting scheme that led to a beam size on the $^{12}$CO(2-1) line cube of 0.06$''$$\times$0.032$''$ 
at an angle of -72$^o$. While we experimented with several spectral bin sizes, the final cube had a resolution of 20 km/s. The final RMS of the cube is $\sim$0.28 mJy/beam. From this cube, we created the moment 0 map described in more detail in the following section, by integrating the line 
emission between -500 km/s and 980 km/s. The RMS of this moment 0 map is 0.04 Jy/beam km/s. For the continuum map discussed in \S~\ref{sec:continuum}, the resulting beam size is 0.078$''$$\times$0.047$''$ at an angle of -73$^o$, while the RMS of the final image is 0.024 mJy/beam.
From the $^{12}$CO(2-1) line cube we also created moment 1 (velocity) and moment 2 (velocity dispersion) maps. The {\it immoments} CASA task was used for this purpose. A 3-$\sigma$ threshold on the total flux was used to create the moment 1 map, while a 5-$\sigma$ limit was 
considered for the velocity dispersion image. These choices are mostly aesthetic and do not have a significant impact on our analysis.





\end{document}